\documentclass[12pt,preprint]{aastex}

\slugcomment{}
\shorttitle{Alpha Persei Corona}
\shortauthors{T.\ R.\ Ayres}
\begin{document}

\title{A Closer Look at the Alpha Persei Coronal Conundrum}

\author{Thomas R.\ Ayres}

\affil{Center for Astrophysics and Space Astronomy,\\
389~UCB, University of Colorado,
Boulder, CO 80309;\\ Thomas.Ayres@Colorado.edu}

\begin{abstract}

A {\em ROSAT}\/ survey of the Alpha Per open cluster in 1993 detected its brightest star, mid-F supergiant $\alpha$~Persei: the X-ray luminosity and spectral hardness were similar to coronally active late-type dwarf members.  Later, in 2010, a {\em Hubble}\/ Cosmic Origins Spectrograph SNAPshot of $\alpha$~Per found far-ultraviolet coronal proxy \ion{Si}{4} unexpectedly weak.  This, and a suspicious offset of the {\em ROSAT}\/ source, suggested that a late-type companion might be responsible for the X-rays.  Recently, a multi-faceted program tested that premise.  Groundbased optical coronography, and near-UV imaging with {\em HST}\/ Wide Field Camera 3, searched for any close-in faint candidate coronal objects, but without success.  Then, a {\em Chandra}\/ pointing found the X-ray source single and coincident with the bright star.  Significantly, the \ion{Si}{4} emissions of $\alpha$~Per, in a deeper FUV spectrum collected by {\em HST}\/ COS as part of the joint program, aligned well with chromospheric atomic oxygen (which must be intrinsic to the luminous star), within the context of cooler late-F and early-G supergiants, including Cepheid variables.  This pointed to the X-rays as the fundamental anomaly.  The over-luminous X-rays still support the case for a hyperactive dwarf secondary, albeit now spatially unresolved.  However, an alternative is that $\alpha$~Per represents a novel class of coronal source.  Resolving the first possibility now has become more difficult, because the easy solution -- a well separated companion -- has been eliminated.  Testing the other possibility will require a broader high-energy census of the early-F supergiants.

\end{abstract}


\section{INTRODUCTION}

Alpha Persei is a luminous mid-F supergiant anchoring the young ($\sim$50~Myr) cluster of that name\footnote{In what follows, ``$\alpha$~Per'' normally will refer to the star, while ``Alpha Per'' normally will denote the cluster.  In cases where ``Alpha Persei'' appears, the specific usage is guided by the context.}.  The bright star was detected in X-rays as a hard coronal source ($T\sim 10$~MK) in a 1993 pointing by {\em ROSAT,}\/ at an X-ray luminosity ($L_{\rm X}$) similar to other cluster members (all late-type dwarfs, except $\alpha$~Per itself: Prosser et al.\ 1996).  However, a subsequent far-ultraviolet (FUV) spectral SNAPshot\footnote{Brief, partial-orbit observations to fill gaps in the {\em HST}\/ schedule.} in 2010 by {\em Hubble's}\/ Cosmic Origin Spectrograph (COS) uncovered what appeared to be a contradictory face of the supergiant (Ayres 2011).  Expected strong coronal-proxy emission from \ion{Si}{4} 1393~\AA\ ($T\sim 8{\times}10^4$~K) was all but absent.  Instead, the FUV spectrum was dominated by a bright continuum of photospheric origin, with numerous superimposed absorption lines, and a few chromospheric ($T\lesssim 10^4$~K) emissions, especially semi-forbidden atomic oxygen at 1355~\AA\ (likely a recombination transition under supergiant conditions).  Some \ion{Si}{4} emission was present, but barely detected and with a narrow line shape, quite unlike the broad subcoronal profiles of typical G-type supergiants.  A closer examination of the {\em ROSAT}\/ $\alpha$~Per pointing, astrometrically registered to optical counterparts from the US Naval Observatory A2.0 Catalog (Monet 1998), suggested that the X-ray centroid was displaced slightly from the coordinates of the bright star, by $\sim 9\arcsec$ (albeit a fraction of the instrumental beam diameter), at a moderate level of significance (Ayres 2011).  

Such offsets had been seen before.  A good example is yellow supergiant $\beta$~Aquarii (G0~Ib).  Reimers et al.\ (1996) originally identified a {\em ROSAT}\/ detection close to $\beta$~Aqr as the bright star, but the association later was questioned by Ayres (2005), who pointed to an apparent $\sim 21\arcsec$ positional discrepancy.  Subsequently, {\em Chandra}\/ resolved the {\em ROSAT}\/ source into a dominant unrelated coronal object, $24\arcsec$ from $\beta$~Aqr, but also a fainter source coincident with the supergiant (Ayres et al.\ 2005).  

To be fair, the {\em ROSAT}\/ displacements in the previous cases described by Ayres (2005) were larger and more significant than for $\alpha$~Per.  Further, in the {\em ROSAT}\/ Alpha~Per survey pointing, the supergiant was 19$\arcmin$ off the boresight, near the shadow of the inner ring of the PSPC\footnote{Position Sensitive Proportional Counter, the main X-ray imager of {\em ROSAT.}} window support ribs, where the imaging is not as good and systematic effects due to the spacecraft dither pattern could come into play.  

Ignoring these caveats for the moment, the suspicious {\em ROSAT}\/ offset, bolstered by the absence of strong coronal-proxy emissions in the COS SNAPshot, raised the possibility that the hard coronal source identified as  $\alpha$~Per might in fact be a close-by hyperactive dwarf cluster member (recalling that coronal activity is heightened in stellar youth [e.g., Stern et al.\ 1981]).  A dwarf companion within $10\arcsec$ of $\alpha$~Per easily could have been missed historically, because an 11th magnitude G star would be lost in the glare of the 2nd magnitude F-type primary.  There was a precedent: another anomalous supergiant coronal source -- $\alpha$~Trianguli Australis ($\alpha$~TrA: K2~Ia) -- was found to have a (very) close-in ($\sim 0.4\arcsec$ separation) faint, probably G-type, dwarf companion thanks to high-resolution 1600~\AA\ imaging by {\em HST's}\/ Wide-Field and Planetary Camera 2 (Ayres et al.\ 2007), taking special advantage of the large contrast of a warm, active G-type spectrum at the shorter wavelengths compared to a cool, low-activity K-type supergiant.  

There was a possible weak link, however, in the chain of reasoning implicating a coronally active dwarf companion of $\alpha$~Per.  The one well-observed (i.e., by both {\em HST}\/ and {\em Chandra}\,) F supergiant -- Canopus ($\alpha$~Car: F0~Ib) -- also was anomalous in the same ways as $\alpha$~Per: X-ray bright but with a very low $L_{\rm Si\,IV}/L_{\rm bol}$ index\footnote{Normalization by the bolometric luminosity helps mitigate the twin biases of different stellar sizes and distances, particularly helpful when considering a mixed sample of remote supergiants and nearby dwarfs.  However, the bolometric normalization expects that the narrow-band flux in the numerator is related in some fashion to the overall luminosity of the star, say through convective production of magnetic flux, as described in the text.  If that implicit connection is not in force, then the bolometric normalization is not appropriate.  A possible example are the Classical Cepheids, described later.  An advantage of the bolometric normalization -- which is derived directly from the $V$ magnitude and a color-dependent bolometric correction -- is that it is somewhat less sensitive to less well known parameters, such as distances for absolute luminosities or angular diameters for surface fluxes.  Nevertheless, to counter any possible disadvantages of the bolometric normalization, absolute luminosities, i.e., $L_{\rm\scriptsize X}$, also are considered in parallel.} (albeit not as low as $\alpha$~Per) compared to early-G supergiants of similar $L_{\rm X}/L_{\rm bol}$.  The Canopus FUV spectrum likewise was dominated by a bright photospheric continuum, superimposed absorptions, and narrow chromospheric emissions, again especially the \ion{O}{1}] 1355~\AA\ intercombination transition.  Canopus had been observed with the {\em Chandra}\/ High Energy Transmission Grating Spectrometer (HETGS) in July 2000.  The zeroth-order spatial map showed the point source falling at the stellar coordinates.  Given the excellent resolution and aspect reconstruction of {\em Chandra,}\/ there can be no doubt that Canopus lacks a visual X-ray companion (i.e., outside $\sim1\arcsec$ from the bright star), although the coronal imaging does not exclude a very close, unresolved secondary.  

The Canopus HETGS spectrum also was hot ($\sim 10$~MK), similar to the hard {\em ROSAT}\/ pulse-height distribution of $\alpha$~Per, and like those of very active, young dwarf stars; but at odds with the depressed $L_{\rm Si\,IV}/L_{\rm bol}$ index (low activity dwarfs, at least, have softer X-ray spectra than their high-activity counterparts [e.g., Dorren et al.\ 1995]).  Together, $\alpha$~Per and Canopus presented a dual coronal conundrum.  Either both harbor unrecognized close hyperactive dwarf companions; or, the two luminous F stars are members of a new class of coronal source with much lower $L_{\rm Si\,IV}/L_{\rm X}$ ratios than the $\sim 1000$~K cooler early-G supergiants.  

To be sure, the early-G supergiants themselves are anomalous compared to G-type dwarfs in an $L_{\rm X}/L_{\rm bol}$ versus $L_{\rm Si\,IV}/L_{\rm bol}$ ``flux-flux'' diagram, showing a systematic displacement to the right of the dwarf-star power-law trend, in the sense of ${\sim}10$~times weaker X-ray emissions at comparable $L_{\rm Si\,IV}/L_{\rm bol}$ ratios (Ayres et al.\ 2005; and as will be illustrated later).  The G supergiants are not exclusively odd, however: fast-rotating late-F/early-G Hertzsprung-gap giants also occupy the same ``X-ray deficient'' locus in the X-ray/\ion{Si}{4} flux-flux diagram (Ayres et al.\ 1998).

There are good reasons to believe that early F-type supergiants might follow a different coronal path than cooler Main sequence stars.  The F supergiants are descendants of massive O or B-type dwarfs, and fall in the upper middle of the Hertzsprung-Russell Diagram, just to the left of the Cepheid Instability Strip, and just at the right edge of the boundary where surface convection ceases to be an important energy transport process in the stellar outer envelope.  Atmospheric densities of such stars are very low, decreasing the heat capacity of the gas and increasing its transparency, both favoring radiation transport over convection.  In solar-like dwarfs, conversely, high densities and high opacities favor kinematic transport processes over radiation, and a thick outer convective envelope is the norm.  

The significance is that coronal activity in cool stars derives from strong surface magnetic fields, which in turn are thought to be produced at the base of the stellar convection zone via an enigmatic ``Dynamo'' (Parker 1970).  The latter is powered by convection, but apparently is strongly catalyzed by stellar rotation; so that young, fast-rotating dwarfs, or tidally synchronized close binaries, fall at the top of the coronal activity heap.  But, among the F-type supergiants with weak convection and long rotation periods, other possibilities for magnetic field production could come into play, and might operate in such a way that the hot, X-ray side of the coronal energy balance is strongly emphasized over the subcoronal component.  In solar-like stars, the coronal-proxy emissions ($T\sim 10^5$~K) are thought to be dominated by cooling of an intermittently heated corona, either directly in terms of gas draining from coronal magnetic loops following cessation of heating, or via electron conduction from hot coronal structures to the lower temperature gas in the underlying $\sim 10^4$~K chromosphere.  If, instead, the coronal heating in early-F supergiants is more sustained in environments where cooling flows are prevented, or conductive heat redistribution is suppressed, the $L_{\rm Si\,IV}/L_{\rm X}$ ratios conceivably could become skewed toward the high-energy side.

Charting the occurrence and properties of hot coronae in the cool half of the H--R diagram is a still unfinished business, despite already four decades of cosmic X-ray exploration.  This especially is true on the warm side, with the F supergiants described above as an example, but also the surprising A-type coronal sources, such as $\alpha$~Aquilae (Altair: A7~V) and $\alpha$~Cephei (A8~V) (Simon et al.\ 1995); and on the cool side with the possibly ``buried coronae'' of the low-activity red giants like Arcturus ($\alpha$~Bootis; K0~III: Ayres et al.\ 2003a), as well as the fully convective late-M stars, where coronae seem to be exclusively flare-dominated, when detected at all (e.g., Fleming et al.\ 2000).  

Arguably, the $\alpha$~Per case was a key piece in that unfinished coronal puzzle, especially if the {\em ROSAT}\/ offset could be confirmed, and the anomalous behavior simply could be ascribed to a companion, rather than invoking exotic new physical processes.  Thus ensued a broad imaging program -- in the optical, UV, and X-rays -- to establish whether the $\alpha$~Per coronal source truly was the supergiant; but, if not, to detect and characterize any close companion at these wavelengths, as well as attempt to record whatever faint X-ray emission might be associated with the supergiant itself (as in the earlier example of $\beta$~Aqr).  A parallel effort was instigated to collect deeper, broader FUV coverage of $\alpha$~Per with {\em HST}\/ COS, especially below 1300~\AA, forbidden to the original SNAPshot survey for detector safety reasons (forced by the then unknown \ion{H}{1} Ly$\alpha$ 1215~\AA\ intensity).  As in the example of the Canopus FUV spectrum, recorded by {\em HST's}\/ Space Telescope Imaging Spectrograph (STIS) in June 2002, the rapidly falling F-type photospheric continuum toward shorter wavelengths should allow key coronal proxies, such as \ion{Si}{3} 1206~\AA\ ($6{\times}10^4$~K) and \ion{N}{5} 1238~\AA\ ($2{\times}10^5$~K), to be captured with minimal interference compared with, say, \ion{Si}{4} 1393~\AA\ at the longer, continuum-swamped wavelengths.

The program will be described more-or-less as it unfolded chronologically.  However, to avoid any possible suspense, the multi-faceted observations failed to confirm a close visual (i.e., resolved) companion to $\alpha$~Per.  On the surface, this supports the alternative idea that $\alpha$~Per, and related Canopus, constitute a novel class of coronal emitter.  Even so, the possibility of {\em unresolved}\/ close companions to both $\alpha$~Per and Canopus still remains open, albeit now more difficult to prove.

\begin{figure}[ht]
\figurenum{1}
\vskip  0mm
\hskip  15mm
\includegraphics[scale=0.75,angle=0]{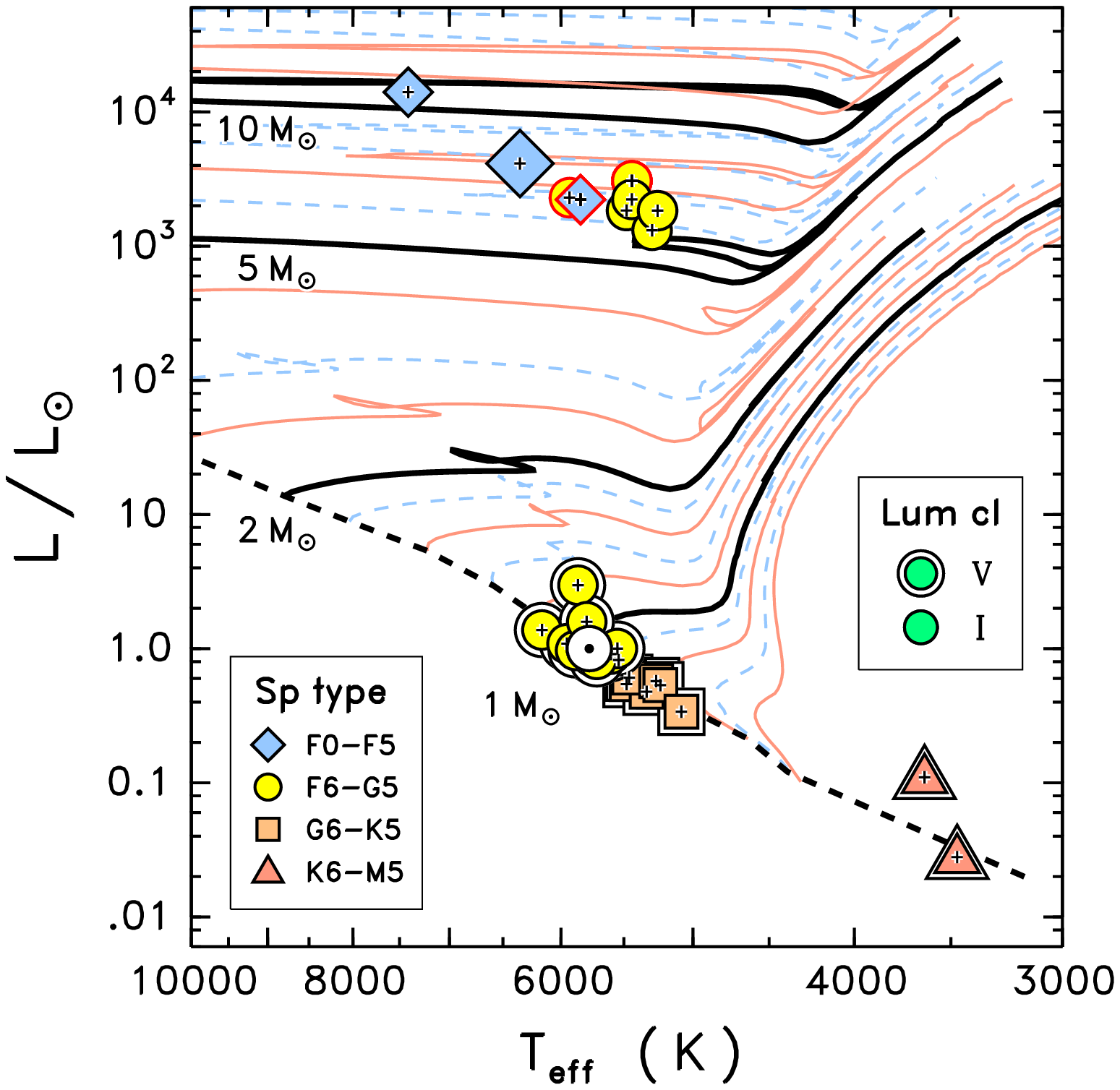} 
\vskip -5mm
\figcaption[]{\small
H--R diagram for G--M dwarfs and F and G supergiants, subjects of various comparisons described later.  Stellar types are marked by symbols and colors according to the two keys.  Location of the Sun is marked $\odot$.  Solid curves are Padova evolutionary tracks (purely illustrative).  Dashed curve is the Zero Age Main Sequence (ZAMS).  Selected tracks (dark, thicker curves) are marked with the ZAMS masses.  Three red-outlined symbols (upper middle) are Classical Cepheids.  Central character of the present study, $\alpha$~Per, is the larger black-bordered diamond in the middle; comparison supergiant Canopus is the leftmost smaller diamond.  
}
\end{figure}

\section{OBSERVATIONS}

\subsection{Alpha~Persei, Canopus, and Comparison Stars}

Before describing the new $\alpha$~Per program, the properties of the supergiant, related object Canopus, and a collection of comparison stars will be summarized.  (The comparison stars will be utilized as a context in a series of ``flux-flux'' diagrams -- pairing off combinations of coronal, subcoronal, and chromospheric diagnostics -- introduced later.)

The initial study (Ayres 2011) already presented a discussion of the stellar parameters of $\alpha$~Per and Canopus, but a few details can be updated.  As reported then, $\alpha$~Per is a non-variable supergiant, although close to the Instability Strip, with a (cluster) age of 50~Myr, mass of $\sim 7\,M_{\odot}$, radius $\sim 63\,R_{\odot}$, and effective temperature $T_{\rm eff}\sim 6270$~K (about 500~K hotter than the Sun).  The spectral type listed in SIMBAD was F5~Ib, and the distance quoted in the earlier study was that of Alpha Per, 190~pc ($191{\pm}7$~pc based on a re-analysis of {\em Hipparcos}\/ open cluster measurements by Robichon et al.\ [1999]).  Note, however, that the {\em Hipparcos}\/ parallax listed in SIMBAD implies a closer distance, $155{\pm}4$~pc, for the supergiant itself; while the multi-color MS fitting approach of Pinsonneault et al.\ (1998) found an intermediate value, $176{\pm}5$~pc, for the cluster center.  More recently, van~Leeuwen (2009) proposed a revised distance of $172{\pm}{3}$~pc for Alpha Per, based on a new analysis of the {\em Hipparcos}\/ material.  Perhaps significantly, the kinematic study by Makarov (2006) placed $\alpha$~Per about 18~pc to the near side of the cluster center.  For the van~Leeuwen cluster distance, this would put the supergiant at about the SIMBAD 155~pc.  The latter value was adopted here.  Makarov (2006) further obtained a cluster age of 52~Myr, by isochrone fitting, on the low side of the more recent estimate by Silaj \& Landstreet (2014), $60{\pm}7$~Myr.

Canopus was reported to be somewhat younger than $\alpha$~Per, at 40~Myr, with SIMBAD spectral type F0~Ib-II, somewhat more massive ($\sim 8\,M_{\odot}$), slightly larger ($R\sim 70\,R_{\odot}$), significantly hotter ($T_{\rm eff}\sim 7560$~K), and closer (96~pc, the {\em Hipparcos}\/ distance).  Canopus currently is listed in SIMBAD as spectral type A9~II, traced back to Gray \& Garrison (1989).  However, in the compilation by Bersier (1996), Canopus is assigned F0~Ib (F5~Ib for $\alpha$~Per), and the Bersier spectral types are consistent with the SIMBAD entries for all the F and G supergiants included in the present study (except, of course, Canopus).  The F0~Ib classification for Canopus was adopted here, although in truth the one sub-type earlier assignment by Gray \& Garrison (1989) is only minimally different.

Figure~1 is an H--R diagram for $\alpha$~Per, Canopus, and about two dozen comparison stars; the latter restricted to G, K, and M dwarfs and F and G supergiants.  Selected Padova evolutionary tracks (e.g., Bertelli et al.\ [2009]: $Z= 0.017$, $Y= 0.23$) are depicted.  These are meant to be purely illustrative, since the specific trajectory for any given star, particularly the more massive ones, depends on details of metallicity, rotational evolution, mass-loss history, and convection treatment.  The fundamental stellar parameters ($L_{\rm bol}$, $T_{\rm eff}$) were derived from a variety of sources, as outlined in Appendix A.  Note that the consensus effective temperatures of $\alpha$~Per and Canopus (derived from the PASTEL catalog of Soubiran et al.\ [2016]) are slightly different than quoted above: 80~K hotter and 150~K cooler, respectively.  One key point of the comparison is that the F and G supergiants have similar temperatures to G and K dwarfs on the MS, but their bolometric luminosities, of course, are several orders of magnitude higher (and their surface gravities correspondingly reduced).  Another key point is that $\alpha$~Per has fundamental properties roughly similar to the non-variable late-F/early-G supergiants, as well as the Classical Cepheids, whereas comparison star Canopus is more separated from the others, especially in temperature.

\subsection{Optical, UV, and X-ray Measurements of Alpha Persei and Environs}

A series of multi-spectral measurements were carried out on $\alpha$~Per, and its immediate vicinity, on the one hand to search for the putative companion suggested by the {\em ROSAT}\/ offset, but on the other to more completely characterize the mid-F supergiant, especially if it was detected as a subsidiary source, or turned out to be the dominant X-ray emitter (as, in fact, was the case).  Table~1 summarizes the various observations described in more detail below.

\begin{deluxetable}{cccccc}
\tabletypesize{\footnotesize}
\tablenum{1}
\tablecaption{Multi-Wavelength Observations of $\alpha$~Persei}
\tablecolumns{6}
\tablewidth{0pt}
\tablehead{\colhead{Dataset} & \colhead{UT Start} & \colhead{$t_{\rm exp}$} & 
\colhead{Splits}  & \colhead{Aperture}  &  \colhead{Filter} \\
\colhead{(1)} & \colhead{(2)} & \colhead{(3)} & \colhead{(4)} & \colhead{(5)}  & \colhead{(6)} 
} 
\startdata
\cutinhead{APO GIFS}
ut150121.10  &  2015-01-21.086  &   500  &  1  &   $-90\degr$  &  York6569 \\ 
ut150121.11  &  2015-01-21.093  &   500  &  1  &   $-90\degr$  &  York6569 \\
ut150121.12  &  2015-01-21.099  &   500  &  1  &   $-90\degr$  &  York6569 \\
ut150121.23  &  2015-01-21.124  &   500  &  1  &  $-135\degr$  &  York6569 \\
ut150121.24  &  2015-01-21.130  &   500  &  1  &  $-135\degr$  &  York6569 \\
ut150121.25  &  2015-01-21.136  &   500  &  1  &  $-135\degr$  &  York6569 \\
\cutinhead{{\em HST}\/ WFC3}
id0512011    &  2015-11-18.346  &   322  &  2   &  UVIS2-2K2C  &  F280N \\
id0512061    &  2015-11-18.411  &   322  &  4   &  UVIS2-C1K1C &  F280N \\
id0513011    &  2016-02-05.632  &   322  &  2   &  UVIS2-2K2C  &  F280N \\
id0513061    &  2016-02-05.692  &   322  &  4   &  UVIS2-C1K1C &  F280N \\
\cutinhead{{\em HST}\/ COS}
ld0510010    &  2015-11-20.209  &  1836  &  4   &     PSA      &  G130M--1291 \\
ld0511010    &  2016-01-07.314  &  1836  &  4   &     PSA      &  G130M--1309 \\
\cutinhead{{\em Chandra}\/ HRC-I}
ObsID\,17723        & 2015-12-22.711   &  20988 &  1   &    $30\arcmin{\times}30\arcmin$  &   NONE \\ 
\enddata
\tablecomments{Col.~3 is total exposure (s), corrected for dead time for HRC-I.  Col.~4 ``Splits'' are cosmic-ray sub-exposures for WFC3 and FP-POS steps for COS, with duration $t_{\rm exp}/N_{\rm split}$.  Col.~5 ``Aperture'' is occulter rotation for GIFS; camera sub-arrays for WFC3; and $2.5\arcsec$-diameter Primary Science Aperture (PSA) for COS.  Col.~6 lists narrow-band filters for GIFS and WFC3, and grating--$\lambda_{\rm cen}$(\AA) for COS.}
\end{deluxetable}

\subsubsection{Ground-Based Optical and Infrared Stellar Catalogs}

Figure~2a illustrates the astrometric reality near bright stars.  USNO~A2 optical catalog entries around $\alpha$~Per (central red circle) are depicted as black circles, and 2MASS (Skrutskie et al.\ 2006) near-infrared sources by red circles.  The symbol size is related to the object magnitude (except $\alpha$~Per): $V\leq15$ for A2 sources, $J\leq15$ for 2MASS.  The brightest sources (largest circles) are about 8th magnitude.  Objects with nearly equal black and red circles are hotter, while those with black circles smaller than the red are cooler.  Both ground-based catalogs have large voids close to the bright star owing to saturation of the older sky survey photographic plates (A2) or the more modern IR digital images (2MASS), and interference from the telescope spider diffraction patterns.  The larger dashed circle is 5$\arcmin$ in radius, about the inner limit of the A2 sources.  The smaller dashed circle is 1.5$\arcmin$ in radius, about the inner boundary of 2MASS.  In order to recover faint, but potentially coronally active, objects closer to the bright star requires a more narrowly focused observational strategy, which in one way or another overcomes the scattered light from the bright central source. 

\clearpage
\begin{figure}[ht!]
\figurenum{2}
\vskip -10mm
\hskip  -10mm
\includegraphics[scale=0.625,angle=0]{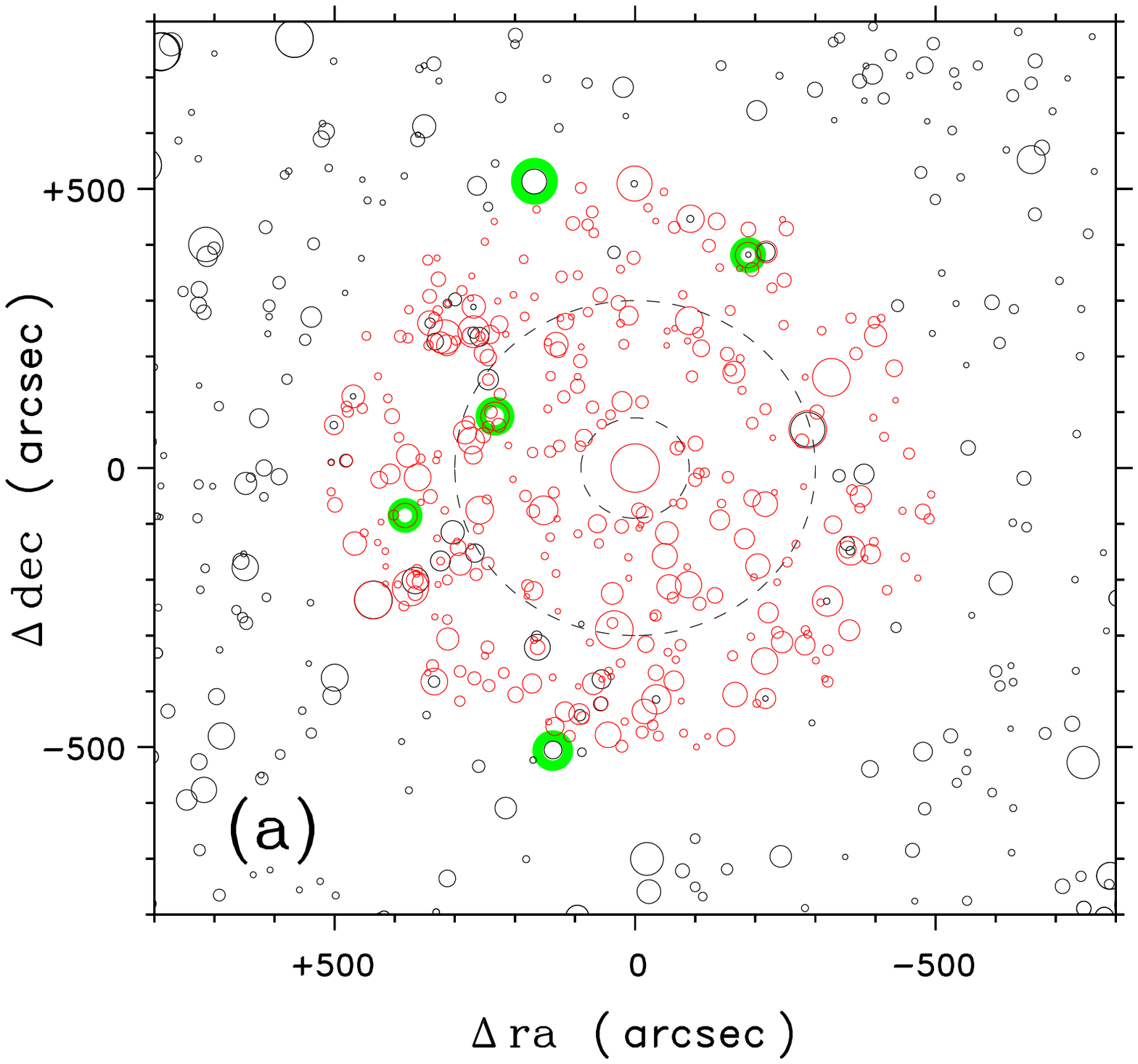} 
\vskip  5mm
\hskip  -11.75mm
\includegraphics[scale=0.525,angle=0]{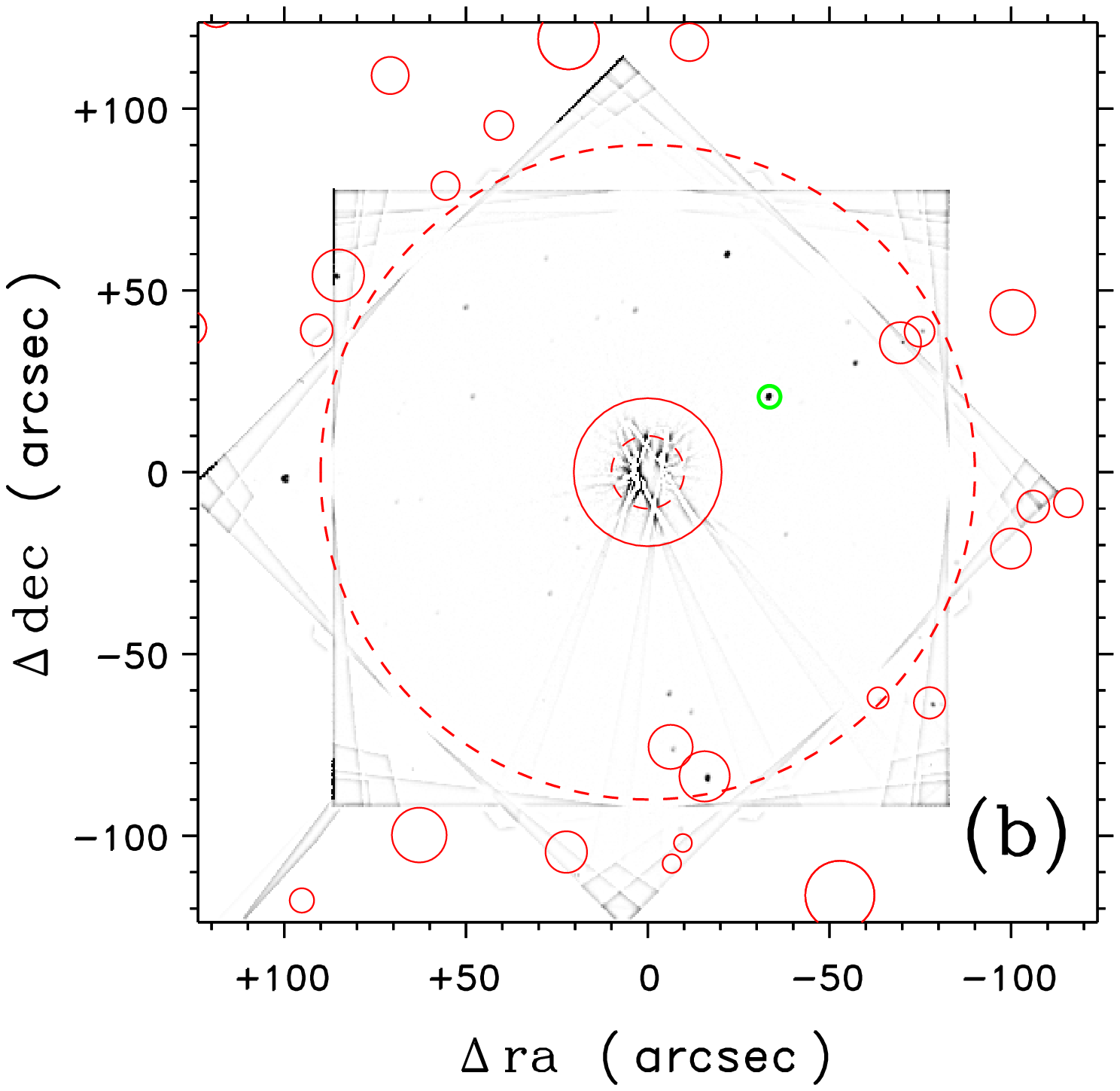} 
\vskip  -95mm
\hskip  76mm
\includegraphics[scale=0.575,angle=0]{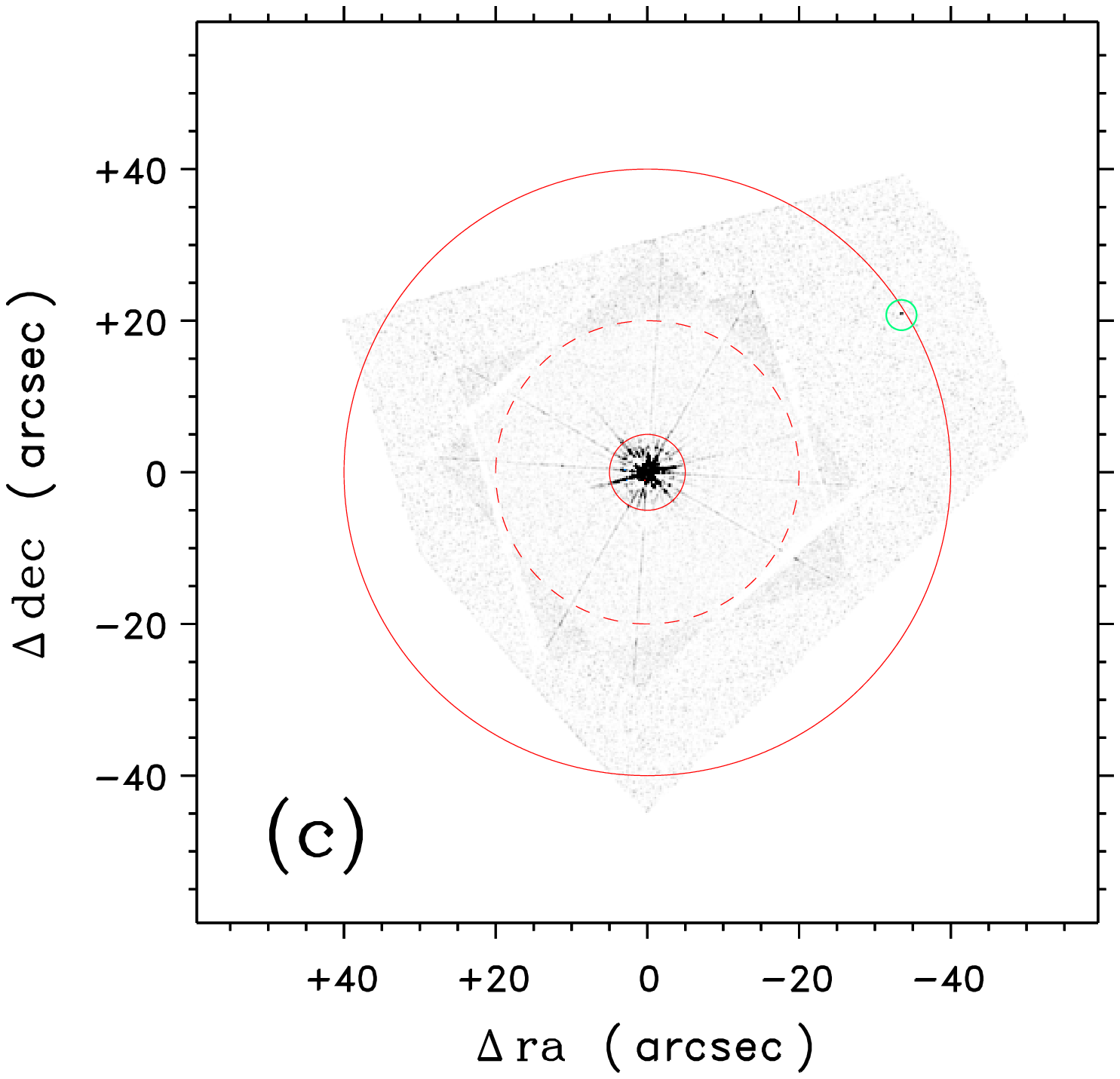} 
\vskip  10mm
\figcaption[]{}
\end{figure}

\clearpage
\begin{figure}[ht!]
\figurenum{2}
\figcaption[]{\small (a) Field around the bright star $\alpha$~Per as portrayed in two catalogs: optical USNO A2 (black circles) and infrared 2MASS (red circles).  Size of the symbol is related to the $V$ magnitude (A2) or $J$ magnitude (2MASS): smallest symbols are $15$th magnitude, largest are $\sim 8$th magnitude (excluding $\alpha$~Per, itself, which is 2nd magnitude in $V$).  Both catalogs have central voids owing to saturation of the original survey material (less extreme for digital 2MASS compared to photographic A2).  The larger dashed circle, $r=5\arcmin$, marks the inner extent of A2; the smaller dashed circle, $r=1.5\arcmin$, the inner boundary of the 2MASS catalog.  (Note: the 2MASS entries are depicted only out to about $9\arcmin$ from the bright star.)  Green-circled objects are optical counterparts of strong {\em Chandra}\/ sources, described later.  (b) Inner $4\arcmin{\times}4\arcmin$ of the $\alpha$~Per field, centered on the bright star, as recorded by the ground-based APO/GIFS coronagraph using a narrow-band H$\alpha$ filter.  Center of the image is dominated by residual artifacts due to incomplete removal of the telescope diffraction pattern and the GIFS occulting finger.  Red circles again represent 2MASS catalog entries: several of the inner ones match point sources seen at the outer peripheries of the GIFS image.  The brightest of these are about $J=14$.  The red dashed circle is $r=1.5\arcmin$.  The green-circled object matches a detection in the WFC3 composite image.  (c) The WFC3 combined image, in a near-UV filter (\ion{Mg}{2} 2800~\AA) centered on $\alpha$~Per.  The several faint radial spokes are artifacts of incomplete removal of the spider diffraction spikes.  The outer solid red circle is $r=40\arcsec$; the dashed red circle is $r=20\arcsec$; and the innermost red circle is $r=5\arcsec$.  There is only one convincing coincidence (green circle) between a point source in the WFC3 image, and a counterpart in the GIFS inner region.  This object is about 14th magnitude, if similar in type to the other stars recorded jointly by GIFS and 2MASS.  There is no sign of the expected much brighter ($V\sim 11$) late-type companion invoked to explain the apparent slight offset of the {\em ROSAT}\/ X-ray source at $\alpha$~Per.
}
\end{figure}

\clearpage
\subsubsection{Optical Coronography from the Ground: APO 3.5-m and GIFS}

An initial attempt to resolve a close-in companion of $\alpha$~Per used the Apache Point Observatory 3.5-m telescope and its Goddard imager and Integral Field Spectrograph (GIFS), an instrument designed mainly to spectrally map circumstellar debris disks, and related structures, around nearby stars.  GIFS employs an occulting finger to block the bright central object.  A complete description of the instrument can be found in Bonfield et al.\ (2008).  

An exploratory program was carried out with GIFS on 2015 January~21 to map the near-field of $\alpha$~Per, during a few hours of engineering time.  The instrument had not been used previously on an object brighter than about 8th magnitude -- one reason for the engineering test -- but the occulting system successfully suppressed $V\sim 2$ $\alpha$~Per down to about $r\sim 10\arcsec$ from the star.  A narrow-band H$\alpha$ filter (York6569: 15~\AA\ FWHM) de-emphasized the F supergiant, which has a deep photospheric hydrogen absorption feature, but favored any young coronally active stars, which often have strong chromospheric Balmer emission.  

The nominal field of view of GIFS, in filter imaging mode, is $2.8\arcmin{\times}2.8\arcmin$ square, which overlaps the inner boundary of the 2MASS catalog around $\alpha$~Per.  The sky is stationary in the GIFS detector reference frame during an observation (usually with N along the $+y$ camera axis) and the occulting wedge also has a fixed orientation relative to the detector (about 45\degr\ from the $+y$-axis toward the $+x$-axis).  However, the GIFS FOV can be rotated to place the sky at any desired position angle relative to the occulting finger.  Although the sky is stationary, the diffraction pattern of the secondary mirror support structure rotates in the GIFS frame during the night owing to the alt-azimuth telescope mount.  This motion causes a slight blurring of the spider diffraction pattern in longer observations.  

The $\alpha$~Per field map described below was constructed from a set of six 500~s exposures taken over about 80~minutes, to build up S/N and for cosmic-ray rejection.  The sequence consisted of two 3-exposure segments, separated by about half an hour.  For the first, the GIFS FOV was positioned so that the occulting finger would be in the SSW quadrant, in sky coordinates, roughly orthogonal to the NW companion suggested by {\em ROSAT.}  For the second segment, the GIFS FOV was rotated an additional 45\degr, placing the occulting finger in the SSE quadrant.

Each three-image segment was processed separately as follows.  First, the three individual CCD frames were rotated to align the diffraction spikes, then stacked in that reference frame, averaging the two lowest values of the three at each pixel.  This effectively removes transient cosmic-ray events and any point sources that follow the sky rotation, but retains persistent structure due to the telescope diffraction pattern.  Next, the filtered diffraction image was scaled to each individual GIFS exposure, still in the diffraction coordinates, and subtracted.  Then, a radial intensity profile, assumed azimuthally symmetric, was determined for each residual image by considering the four sectors around the point source, avoiding the radial locations of the diffraction spikes.  The intensities in these sectors were filtered to remove the point sources (stars and cosmic-rays), and azimuthally averaged.  The resulting radial ``scattering profile'' was subtracted from each residual image.  Next, the background-subtracted individual frames were de-rotated back into sky coordinates.  In the resulting three-image stack, the maximum value at each sky pixel was rejected, then the two surviving lesser values were averaged.  The filtering effectively excluded cosmic rays as well as high-signal residual artifacts associated with removal of the background diffraction spikes, while emphasizing features that rotate with the sky and were present in at least two of the time-diverse exposures. 

The two separate segment images, with the different occulting wedge rotations, then were merged, taking the maximum of the two, pixel-by-pixel, beyond $r=20\arcsec$ from the center, but the minimum inside that radius.  This approach retains the positive signals in the outer regions where only one of the segments might contribute (owing to the relative rotation of the two sequences and the square shape of the detector array), but captures the lessor of the two signals in the inner region where the residual diffraction patterns are mixed together.  Finally, a highly filtered version of the image was subtracted to remove any lingering larger scale intensity variations.

The composite GIFS image is illustrated in Figure~2b.  It captures a dozen, or so, previously uncatalogued optical objects inside the roughly $r=90\arcsec$ exclusion zone of 2MASS, but also matches up with a number of the 2MASS point sources at, and outside, that radius.  The two relatively bright objects at the periphery of the merged image (to the S and NE of $\alpha$~Per), which line up with 2MASS entries, have $J$ magnitudes of about 14.  None of the new GIFS objects, however, are as close to $\alpha$~Per as suggested by the {\em ROSAT}\/ offset.  To be sure, the putative companion would be right at the edge of the occulted zone around the bright star, where unfortunately the residual diffraction artifacts are a source of confusion (although an 11th magnitude companion should easily have been seen at the boundary).

There were other, potentially more powerful, ground-based coronographs available for the purpose, like the {\em Gemini}\/ Planet Imager (Macintosh et al.\ 2006) on the {\em Gemini-S}\/ Telescope, or SPHERE (Spectro-Polarimetric High-contrast Exoplanet Research: Beuzit et al.\ 2006) on the European Southern Observatory Very Large Telescope.  Unfortunately, the FOVs were too narrow ($\sim 2.8\arcsec$ on a side for GPI; $\sim 1.8\arcsec$ for SPHERE) to probe the expected larger ($\sim 9\arcsec$) displacement suggested by the {\em ROSAT}\/ offset, and the respective observatories also were located too far south for northern-situated $\alpha$~Per.  So, a different strategy had to be pursued.  The new instrument of choice was {\em HST's}\/ Wide Field Camera 3 (WFC3), which has a $2.7\arcmin{\times}2.7\arcmin$ field of view in its UV/visible channel (``UVIS''), excellent suppression of a central bright star owing to the high spatial resolution of the space-borne telescope, together with the stable point spread function thanks to lack of atmospheric interference.

\subsubsection{Close-in Near-UV Imaging with {\em HST}\/ WFC3-UVIS}

The WFC3 observations of $\alpha$~Per were carried out as part of the joint {\em Chandra/HST}\/ program alluded to earlier.  The purpose of the WFC3 part was to image the near field of the bright star in the visible and near-UV, to ferret out any faint, low-mass dwarf that might fall close to the supergiant, either coincidentally or possibly as a wide binary companion.  The environs around $\alpha$~Per already had been mapped in H$\alpha$ with the GIFS coronagraph, as described earlier, down to $r\sim10\arcsec$.  The innermost region, however, was partially blocked by the GIFS occulting finger and affected by incomplete suppression of stray light from the bright star.  

For this reason, only a subset of the WFC3-UVIS field was needed.  In fact, recording the whole FOV would have been very inefficient owing to the short exposure times involved, in the face of the significant buffer transfer overheads for the full 4K$\times$4K frames.  Instead, context imaging was provided by the UVIS2-2K2C $\frac{1}{4}$-subarray, to reduce the instrumental overheads, but still provide an $80\arcsec{\times}80\arcsec$ FOV, entirely adequate for the purpose (i.e., overlapping the inner part of the GIFS H$\alpha$ map).  Three narrow-band filters -- F280N (\ion{Mg}{2}), F395N (\ion{Ca}{2}), and F656N (H$\alpha$) -- were used.  An active low-mass star would show enhanced chromospheric {\em emission}\/ in these filters, in contrast to the broad {\em absorption}\/ features anticipated for the warm supergiant.  Exposure times were chosen for each filter to maintain about the same level of saturation of the bright star, but also allow an adjacent solar-spectrum companion with $V\sim 11$ (typical of an active Alpha Per G dwarf) to be imaged with S/N$\sim$150.  The context exposures were (CR-)split into pairs for cosmic-ray mitigation.  Post-flashes of 12 e$^{-}$ were imposed to improve the Charge Transfer Efficiency (CTE) in these otherwise low-background exposures (away from the saturated central point source).  

In addition to the three 2K$\times$2K context images, one for each filter, a second series was taken in the UVIS2-C1K1C $\frac{1}{8}$-subarry to focus on the $40\arcsec{\times}40\arcsec$ region close to $\alpha$~Per where an interloping X-ray source was most likely to be found, at least given the suggestive {\em ROSAT}\/ offset.  Use of the $\frac{1}{8}$-subarray forced the buffer transfers into the earth occultation, so that no exposure time was lost.  The same total exposure duration was applied to each of the three filters, except now the observations were CR-split into 4, effectively halving the sub-exposure integrations, and thus also the $\alpha$~Per saturation, while improving cosmic-ray rejection.  In the central region imaged by both the $\frac{1}{4}$-subarray and the $\frac{1}{8}$-subarray, the total coverage effectively was doubled.  This hierarchical scheme was judged the most efficient way to tie the inner $\alpha$~Per field to the existing astrometry provided by the GIFS imaging (itself referenced to the 2MASS system), and achieve a high level of exposure in the central region without losing time to buffer dumps.  Since high-precision astrometry was not required, explicit dithering was not used.  A gross level of dithering, nevertheless, was accomplished by combining the 2K$\times$2K and 1K$\times$1K imaging. 

Two single-orbit WFC3 visits were needed, to shift the telescope and camera artifacts relative to the sky.  These pointings were carried out on 2015~November 18 and 2016~February 5.  An ORIENT constraint was introduced in the first visit to place the putative low-mass active companion (based on the Position Angle [PA] deduced from the {\em ROSAT}\/ offset) in approximately the same CCD row as the bright star, to minimize the influence of bleeding (down the columns: the CCD $y$-axis) from the saturated image of $\alpha$~Per, and avoid the telescope diffraction spikes (a cross pattern located approximately at $\pm$45\degr\ and $\pm$135\degr\ with respect to the detector $y$-axis).  In the second visit, a 30\degr\ rotation clockwise from the optimum orientation was imposed to minimize the effect of diffraction spikes (and bleeding) on the combined field map, especially in the event that any close-in companion star was at a different PA than suggested by the {\em ROSAT}\/ observation.

The post-processing made use of the pipeline {\sf .drz} (``drizzled'') files, which include CR-rejection (in either the 2K$\times$2K pairs, or 1K$\times$1K quartets), then co-addition of the frames, geometrically corrected for image distortions, and finally re-sampled onto a uniform sky grid with 0.0396$\arcsec$ pixels.  The longer duration F280N images showed the best results, at least in terms of revealing objects in the $\alpha$~Per field that also were present in the GIFS H$\alpha$ map.  A straightforward image combination strategy was adopted.  The approach was to consider the four individual F280N exposures -- two UVIS2-2K2C and two UVIS2-C1K1C, with each 2K2C/C1K1C pair at different ORIENTs -- as a coherent set.  Each individual image first was filtered to remove small scale bright spots (mainly residual cosmic rays) and average over large-scale intensity variations.  Then, the smoothed ``background'' was subtracted from that frame.  Next, the background-subtracted frames were rotated into common sky coordinates.  Finally, the stack of four images was filtered so that if two or more positive values were present at a given sky pixel, the minimum value would be selected; otherwise the image pixel was set to zero.  This is an expedient way to suppress the residual telescope (and detector) artifacts, while still retaining any positive signals due to persistent sky structure (such as stars).

The filtered merged F280N map of $\alpha$~Per is illustrated in Figure~2c.  WFC3 clearly has opened up the inner region around the bright star, down to about $r=5\arcsec$, where incomplete removal of the diffraction artifacts dominates the point-source detectability.  There is, in fact, only a single convincing match between an apparent WFC3 point source outside $r=5\arcsec$ and the GIFS detections.  This coincident object was too far from $\alpha$~Per, and too faint, to be a candidate for the {\em ROSAT}\/ offset source; and there were no obvious objects at, or inward of, the 9$\arcsec$ NW position suggested by {\em ROSAT.}\/ 

Thus, the initial conclusion was that the suspicious PSPC offset was in fact illusory.  In hindsight, this was understandable, given the off-boresight location of $\alpha$~Per in the original {\em ROSAT}\/ pointing, with attendant localization issues; but disappointing nonetheless because the easy resolution of the $\alpha$~Per X-ray anomaly -- a close companion as in the previous case of $\beta$~Aqr -- had not been realized.

\subsubsection{{\em HST}\/ COS FUV Spectroscopy}

{\em HST's}\/ Cosmic Origins Spectrograph (COS) was the clear choice to obtain a deep far-ultraviolet (FUV) spectrum of $\alpha$~Per.  Even though the star is bright in the optical, it is too faint in the FUV (as determined from the earlier COS SNAPshot) to be accessible to the other UV instrument on {\em Hubble,} high-resolution STIS.  (On the other hand, STIS was able to successfully record the 1150--1700~\AA\ FUV spectrum of comparison supergiant Canopus, which is about 200 times brighter than $\alpha$~Per at 1400~\AA\ in absolute flux densities.)  COS had the further potential advantage of a modest degree of spatial discrimination within its 2.5$\arcsec$-diameter Primary Science Aperture (PSA), a region inaccessible to the WFC3 imaging owing to saturation of the central point source.

The COS FUV spectroscopy was done in two single-orbit visits, on 2015 November 20 and 2016 January 7.  The ORIENT angles were approximately orthogonal in the independent visits, to allow post-facto separation of a potentially doubled spectrum, in the event that the putative low-mass active companion of $\alpha$~Per happened to fall within an arcsecond, or so, of the bright star.  The main spectral focus of the COS part of the program was the short-FUV interval (1150--1450~\AA) from the G130M grating.  Because of the rapidly rising F-star continuum toward longer wavelengths, which would overwhelm any emission from lines like \ion{C}{4} 1548~\AA, there was no good reason to expend any additional time to obtain, say, a long-FUV (1400--1700~\AA) observation with grating G160M.  The resolution ($\lambda/\Delta\lambda$) of G130M is about 18,000 (17~km s$^{-1}$).

In each visit, $\alpha$~Per was acquired in dispersed FUV light using G130M itself: the star is too bright in the longer wavelength near-UV for a direct-imaging capture with the NUV-MAMA camera.  Initially, a 9-step raster search isolated the target coarsely, then the centering was refined with a PEAKXD (``peak-up'' [centroiding] in the cross-dispersion direction), and lastly by a PEAKD (peak-up along the dispersion direction).  The remainder of the single orbit was occupied by a sequence of four G130M exposures, 459~s each, at the standard FP-POS steps (small grating rotations intended to mitigate fixed-pattern noise).  The total exposure of 1840~s in each visit was quadruple that of the brief G130M SNAP in 2010.  The low activity level of $\alpha$~Per deduced from the earlier COS spectrum allayed any concerns over a bright limit violation at Ly$\alpha$, which allowed detector side B (1150--1290~\AA) to be activated.  (The unknown FUV activity level of $\alpha$~Per prior to the SNAP program originally had prevented use of G130M side B, for detector safety reasons.)  The two new $\alpha$~Per visits were done with separate G130M wavelength settings (``CENWAVE'') -- 1291~\AA\ and 1309~\AA\ -- to avoid the small spectral gap that would have been present if the same CENWAVE had been used in both visits. 

The two independent sets of G130M exposures were processed through the {\sf CALCOS} pipeline, which combined the four separate FP-POS sub-exposures of each visit, screened for fixed pattern defects such as grid wire shadows.  Spatial/spectral images of the two independent G130M observations were constructed directly from the event lists (see, e.g., Ayres 2015a), but there was no evidence for any cross-dispersion asymmetries or spectrum doubling that might indicate the presence of a partially resolved companion.  Time histories of key line fluxes, such as \ion{Si}{3} 1206~\AA, \ion{N}{5} 1238~\AA, and \ion{O}{1} 1302~\AA, also were extracted from the event lists, but no obvious transient behavior (such as a flare) was seen in either, other than the strong influence of atomic oxygen skyglow on the 1305~\AA\ resonance lines in the first visit (with the 1291~\AA\ CENWAVE; the 1309~\AA\ setting has the detector gap on top of the \ion{O}{1} region).  The two independent FUV tracings then were combined.  The \ion{O}{1} region (1300--1309~\AA) was treated separately owing to the skyglow contamination.  For this interval, the 1291~\AA\ sub-exposure with the minimum integrated \ion{O}{1} intensity (out the four FP-POS splits in the first visit) was spliced in, representing, ideally, minimal skyglow distortion of the intrinsic stellar \ion{O}{1} resonance features.

\clearpage
\begin{figure}[ht!]
\figurenum{3}
\vskip  0mm
\hskip  9mm
\includegraphics[scale=0.825,angle=0]{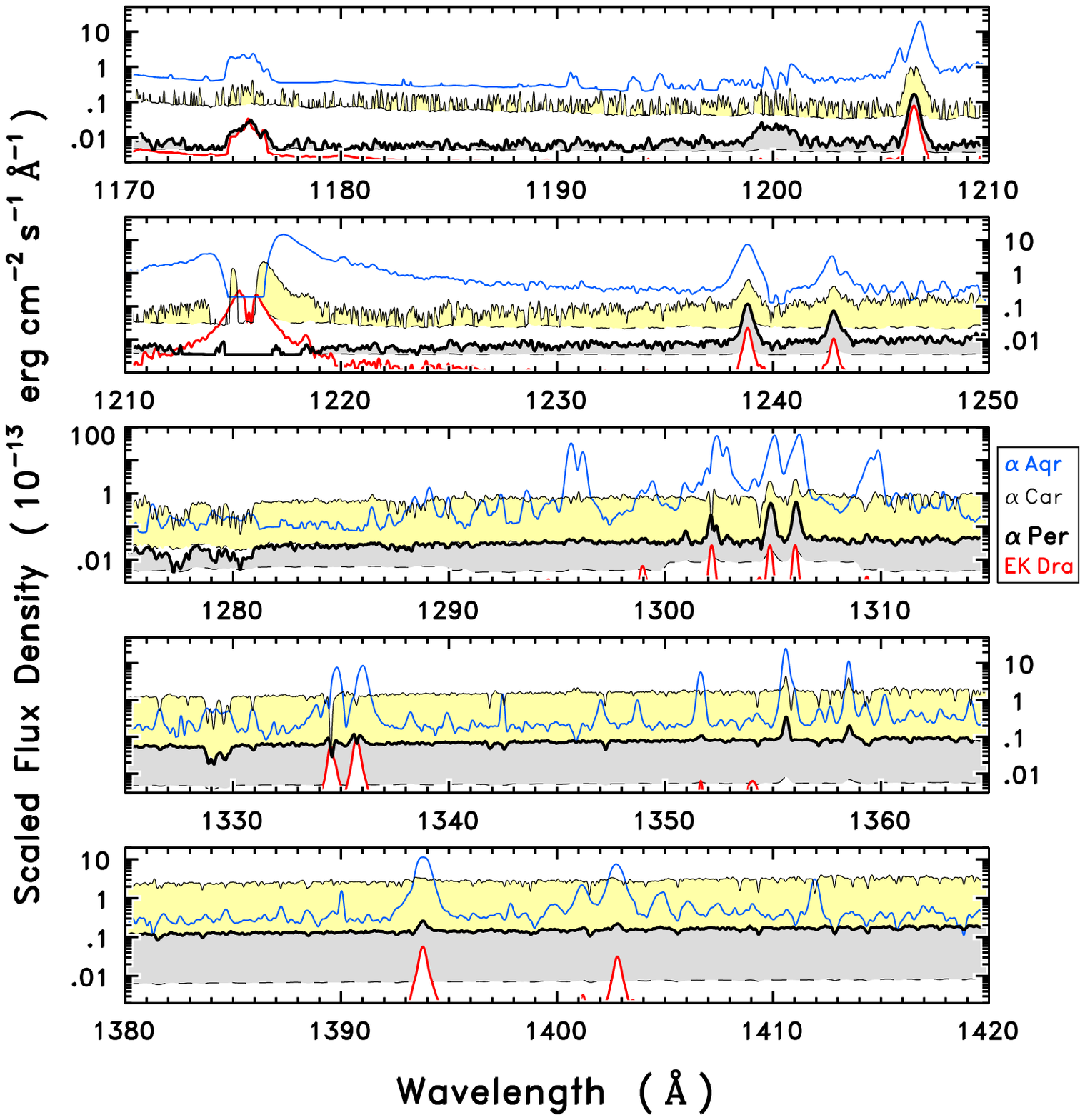} 
\vskip  0mm
\figcaption[]{\small
Gaussian smoothed (FWHM$\sim 30$~km s$^{-1}$) FUV spectra of $\alpha$~Per (thick black outlined, gray shaded) and two other supergiants -- Canopus (F0~Ib: thin black outlined, yellow shaded) and $\alpha$~Aqr (G2~Ib: thin blue curve) -- as well as Alpha-Per-age solar analog EK~Dra (G2~V: red outlined, white shaded).  (See color-coding key to the right of the middle panel.)  The lower edges of the shaded areas, marked by black dashed curves, represent smoothed 1\,$\sigma$ photometric errors (per resel, accounting for the Gaussian filtering).  Note the elevated photometric error of the $\alpha$~Per spectrum at 1300--1309~\AA\ where only one of the sub-exposures was included (selected to minimize {O}\,{\scriptsize I} skyglow contamination).  The flux density scale refers to $\alpha$~Per.  The other supergiants were adjusted according to their optical intensities relative to $\alpha$~Per, while the hyperactive G dwarf was scaled according to $d^2$ relative to $\alpha$~Per.  The flux density curves were truncated at the 1\,$\sigma$ level for this logarithmic presentation.  The flux limiting is more obvious for the shaded tracings, but also affects the single curve for $\alpha$~Aqr, especially at the shortest wavelengths where the sensitivity is falling (and the photometric error rising) rapidly.
}
\end{figure}

\clearpage
\begin{figure}[ht!]
\figurenum{4}
\vskip -5mm
\hskip +15mm
\includegraphics[scale=0.75,angle=0]{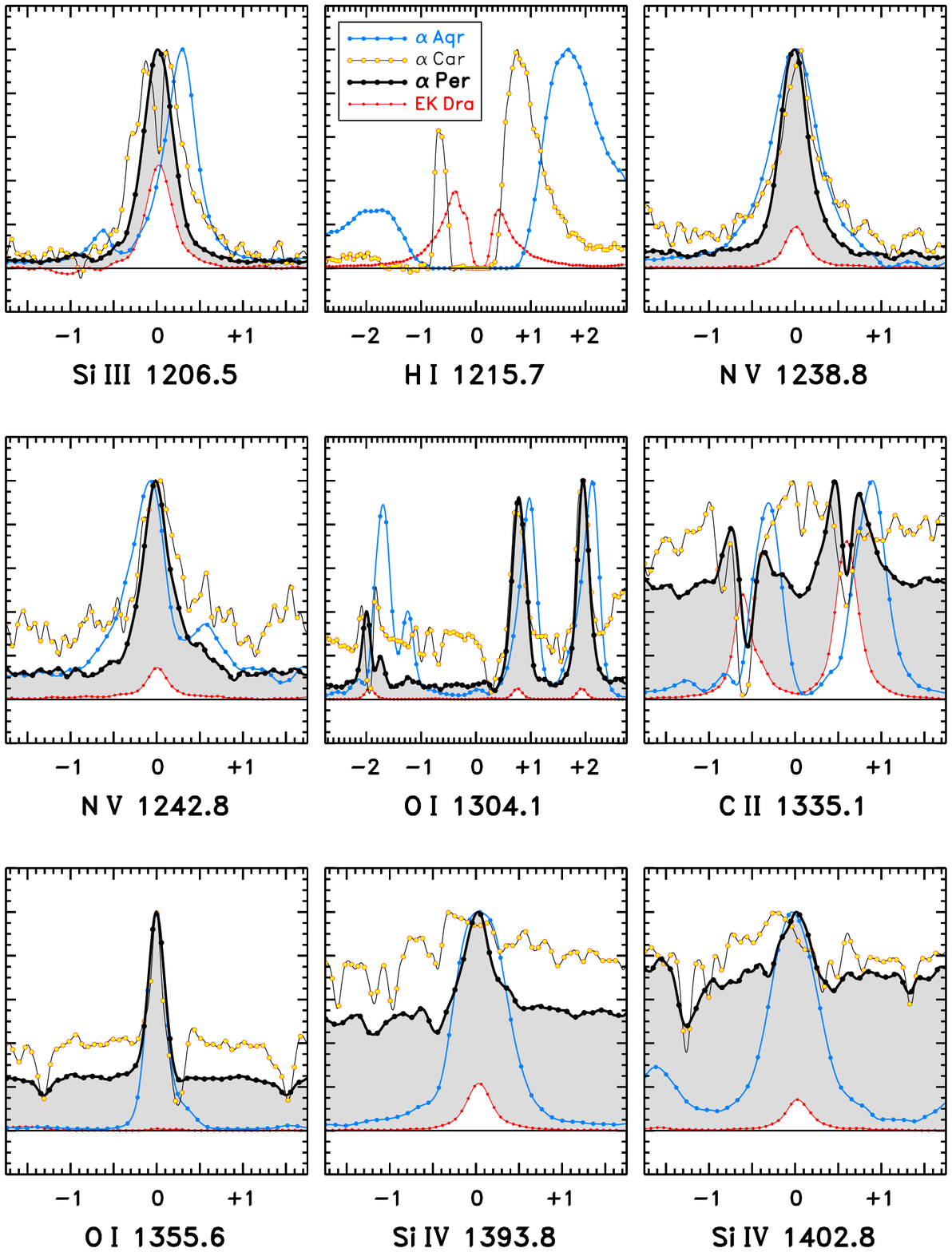} 
\vskip -5mm
\figcaption[]{\small
FUV spectra of $\alpha$~Per and comparison stars, now on a linear flux density scale, and adjusted to the maximum in each interval; except for EK~Dra, which was scaled as $d^{2}$ relative to $\alpha$~Per for all but the Ly$\alpha$ panel ($\alpha$~Per omitted owing to severe contamination by geocoronal {H}\,{\scriptsize I} emission), where the $d^{2}$ is relative to Canopus.  The $x$-axis scale is $\Delta\lambda$ (\AA) relative to the wavelength listed in the axis title.  Color-coding of the spectra is according to the key in the middle upper panel.
}
\end{figure}

The merged COS G130M FUV spectrum of $\alpha$~Per is compared to those of several reference stars in Figure~3, including the STIS spectrum of Canopus mentioned earlier.  Each spectrum was smoothed by a Gaussian filter with a FWHM of approximately 2 COS resolution elements (resel).  The $\alpha$~Per flux densities are displayed on their native scale, while the other supergiants were adjusted according to their relative visual fluxes (${\times}10^{+(V_{\star}-V_{\alpha\,{\rm Per}})/2.5}$).  The COS FUV spectrum of the active solar analog EK~Draconis (G2~V: Ayres 2015a) was scaled according to $(d_{\rm EK\,Dra}/d_{\alpha\,{\rm Per}})^2$, equivalent to a $V\sim 11$th magnitude early G dwarf in the Alpha Per cluster. 

At the longer wavelengths, $\lambda>1380$~\AA, the two warm supergiants -- Canopus and $\alpha$~Per -- display continuum-dominated energy distributions, the former more elevated than the latter owing to $\sim 1000$~K hotter photospheric temperatures.  Nevertheless, the yellow supergiant $\alpha$~Aqr also displays an elevated continuum, even above that of $\alpha$~Per on the $V$--adjusted relative flux density scale, despite its $\sim 1000$~K cooler photosphere.  This is because the G star's chromosphere is (more) active and optically thick enough that the FUV continuum has a significant contribution from the hotter chromospheric temperatures (Planck-weighted exponentially relative to the cooler photosphere).  At the same time, the G supergiant clearly is emission-line dominated (noting the many-decade logarithmic flux density scale) compared with the two F-type supergiants.  The active G dwarf EK Dra also is emission-line dominated; so much so that its hot, chromospheric controlled continuum is not even visible on this relative scale.  It is particularly striking that the \ion{Si}{4} emissions of the two G stars, supergiant and dwarf, are conspicuous, while those of the two F supergiants are barely visible, at least in this broad view.

Note, also, that the two F supergiants have strong ground-configuration \ion{C}{1} multiplet absorptions (e.g., 1276--1280~\AA, 1329--1330~\AA), whereas the same multiplet is in emission in the G supergiant.  The absorption behavior in the F stars is indicative of formation within a photosphere, namely temperatures falling with increasing altitude so that the optically thicker carbon line cores arise at lower temperatures and thus lower intensities than the thinner continuum, which arises deeper in.  Conversely, the emission behavior in the G supergiant points to formation in an atmosphere with temperatures rising outward, namely a chromosphere.  

On the other hand, the \ion{C}{3} multiplet at 1175~\AA\  ($6\times10^4$~K), is in emission in all four objects, at about the same relative levels in $\alpha$~Per and EK~Dra; similar to the \ion{Si}{3} 1206~\AA\ feature ($6\times10^4$~K), but different from most of the other higher-excitation emissions (such as the \ion{N}{5} 1240~\AA\ doublet [$2\times10^5$~K]), which are weaker in EK Dra than in $\alpha$~Per (in this specific relative sense).  Furthermore, the several \ion{O}{1} features, from resonance and intersystem multiplets, are in emission in all three supergiants as well as the active dwarf.  

One conspicuous difference among the stars is \ion{H}{1} Ly$\alpha$, which basically is absent in $\alpha$~Per (to the extent that can be judged given the strong geocoronal contamination of the observed feature), relatively narrow in both Canopus and EK~Dra, but very broad in $\alpha$~Aqr.  The Canopus profile further displays a strong absorption blueward of the presumably mostly interstellar absorption core, which likely indicates a current expanding wind or an archaic shell of material from a prior evolutionary stage (Brown et al.\ 2003).

A closer view of selected spectral features is provided in Figure~4, now on a linear scale.  The flux densities were normalized to the peak values of each star in each panel; except for EK~Dra, which was scaled as $d^2$ (see above) relative to $\alpha$~Per in all the panels except for Ly$\alpha$ (\ion{H}{1} 1215~\AA), where the $d^2$ scaling is relative to Canopus ($\alpha$~Per was omitted from this panel because the interval was obliterated by geocoronal atomic hydrogen emission).  Also, the COS profiles of $\alpha$~Per were shifted relative to the STIS spectrum of Canopus by a few~km s$^{-1}$, after accounting for the stellar radial velocities, to align the low-excitation narrow chromospheric emissions of the \ion{O}{1} 1305~\AA\ multiplet (specifically 1304~\AA\ and 1305~\AA, which, unlike companion 1302~\AA, are not affected by ISM absorption) and the \ion{O}{1}] 1355~\AA\ intercombination transition.  The COS wavelength scales are known to suffer from stretching effects (Ayres 2015a), and the zero-point shift compensates to some extent.

These more close-up views of the resolved line profiles reveal additional oddities.  First, consider \ion{O}{1}] 1355~\AA.  This is the one emission feature that apparently has the same (narrow) shape in all three supergiants (and is so weak, relatively, in the dwarf to be essentially invisible).  In the two F supergiants, especially hotter Canopus, \ion{O}{1}] 1355~\AA\ is flanked by sharp \ion{C}{1} photospheric absorptions (from excited states, 1.26~eV above ground), whereas in the G supergiant, the three \ion{C}{1} features in this interval are in emission (in the case of \ion{O}{1}] 1355~\AA, \ion{C}{1} 1355~\AA\ is the slight red shoulder at the base of the profile).  In fact, although not discernible with this scaling, the \ion{C}{1} 1355~\AA\ feature of EK~Dra is almost the same intensity as neighboring \ion{O}{1}] 1355~\AA\ (see Ayres [2015a], his Fig.~2), which seems to be a characteristic of very active dwarf stars (see Ayres et al.\ [2003b], their Fig.~3).

Although the \ion{O}{1} intersystem line is nearly identical in shape in the three supergiants, the atomic oxygen resonance triplet at 1305~\AA\ displays significant differences, especially between the two F supergiants and the G-type counterpart $\alpha$~Aqr.  The \ion{O}{1} 1304~\AA\ and 1305~\AA\ components of $\alpha$~Per and Canopus have similar, narrow profiles more-or-less centered on the rest wavelengths (as are the much weaker, again relatively speaking, emissions of the G dwarf EK Dra).  Conversely, the $\alpha$~Aqr features appear to be strongly redshifted.  In reality, the appearance is caused by strong blueward absorption of an intrinsically broader chromospheric profile by an expanding warm wind.  The ground-state resonance line is more complicated, because on top of any wind absorption components, there usually are narrower dips due to the interstellar medium, which might be at different velocities in the different stars depending on the particular clouds intersected for those lines-of-sight.  In G supergiant $\alpha$~Aqr, there also is a fluoresced atomic sulfur emission just redward of the \ion{O}{1} 1302~\AA\ peak, which renders the apparent line shape even more complex.  

The same caveats apply to the ionized carbon resonance lines at 1335~\AA.  Here, however, the EK~Dra features are more substantial, and basically point to the ``rest'' positions of the lines, modulo any peculiar velocities due to, say, chromospheric upflows or downflows.  The $\alpha$~Aqr \ion{C}{2} features again display strongly redshifted peaks, symptomatic of broad blueshifted absorption due to a strong outflow at those temperatures ($\sim 3\times10^4$~K).  The $\alpha$~Per profiles are somewhat intermediate, showing a self-reversed emission feature on top of the dominantly photospheric continuum, but the origin of the central dip -- wind, photosphere, ISM, or a combination -- is not clear.  The situation for Canopus is even more convoluted.  Any chromospheric \ion{C}{2} emission is obscured by a series of absorptions: a dominant one near the stellar velocity, but also a possible series of subordinate absorptions at blueward velocities, perhaps lining up with the blueshifted circumstellar features seen in Ly$\alpha$. 

Speaking of the hydrogen resonance line, the close-up view in Fig.~4 repeats some of the differences seen in the \ion{O}{1} and \ion{C}{2} counterparts, but also additional complexity, especially the pervasive blueward absorption in the Canopus profile beyond the interstellar core (${\pm}0.3$~\AA), leaving a peculiar small island of emission at $-0.7$~\AA.  The \ion{H}{1} feature of $\alpha$~Aqr is very broad compared to Canopus, and again shows the effects of the supergiant wind.  At the same time, the hydrogen profile of the dwarf EK~Dra is much narrower (Ly$\alpha$ ``Wilson-Bappu Effect'' [cf., Wilson \& Bappu 1957]).  In addition to the central ISM dip of EK~Dra, there is a weaker blueward notch due to interstellar \ion{D}{1} absorption.  Given the appropriate $d^2$ scaling, a G dwarf as active as EK~Dra in orbit around Canopus would have its Ly$\alpha$ emission mostly submerged in the brighter Canopus feature, and diminished by the several absorption effects, especially if the extra blueward dip of the F supergiant is due to a far away circumstellar shell that would attenuate the dwarf's emission as well.

Consider, now, the hot lines \ion{Si}{4} ($T\sim 8\times10^4$~K) and \ion{N}{5} ($2\times10^5$~K).  Both components of the \ion{N}{5} doublet are clearly present in all four stars, somewhat narrower in the two F supergiants compared with G-type $\alpha$~Aqr; and significantly weaker in the (scaled) active dwarf EK~Dra than in $\alpha$~Per.  The 1242~\AA\ component of $\alpha$~Aqr appears to display a blueward absorption that is not present in the stronger component (1238~\AA) and thus must be from an unrelated species, probably \ion{C}{1} given the prevalence of carbon absorptions near one of its excited state ionization limits ($\sim 1240$~\AA).  The same absorption structure might also be present in the, albeit noisier, Canopus profile.

As for \ion{Si}{4}, the doublet features appear convincingly in the $\alpha$~Per spectrum, although clearly affected by the strong underlying continuum; but they are not at all obvious in Canopus.  The peak-flux normalization in Fig.~4 is biased against Canopus, however, which has a much brighter continuum level at 1400~\AA\ (in $f_{\lambda}/f_{V}$), so it would be easier to hide the same strength \ion{Si}{4} emission than in weaker-continuum $\alpha$~Per.  Line-dominated $\alpha$~Aqr displays more Gaussian-shaped, and somewhat broader, profiles of both \ion{Si}{4} components than the sharper, almost triangular line shapes of the \ion{N}{5} counterparts.  (The smaller emission feature blueward of 1402~\AA\ is a member of an intercombination multiplet of \ion{O}{4}.)  The \ion{Si}{4} lines of EK~Dra are narrow like those of $\alpha$~Per (above the latter's continuum), and are perhaps half the strength (with the $d^2$ scaling).

A summary of the FUV segment of the program is that $\alpha$~Per and the two comparison supergiants show remarkably different chromospheric and higher temperature line shapes, in part because the G-type supergiant is affected by a strong, but relatively low-velocity mass outflow ($50-100$~km s$^{-1}$), while F-type Canopus also displays evidence for circumstellar absorption, but at much higher velocities ($>300$~km s$^{-1}$).  On the other hand, mid-F $\alpha$~Per does not appear to be affected by extraneous absorption to the same extent as the other two supergiants, although the obliteration of the Ly$\alpha$ profile by geocoronal emission has removed a valuable clue concerning what circumstellar environment might surround $\alpha$~Per (the \ion{H}{1} resonance absorption normally is much more sensitive to such material than any other species).

\clearpage
\begin{figure}[ht!]
\figurenum{5}
\vskip  -5mm
\hskip -3mm
\includegraphics[scale=0.85,angle=90]{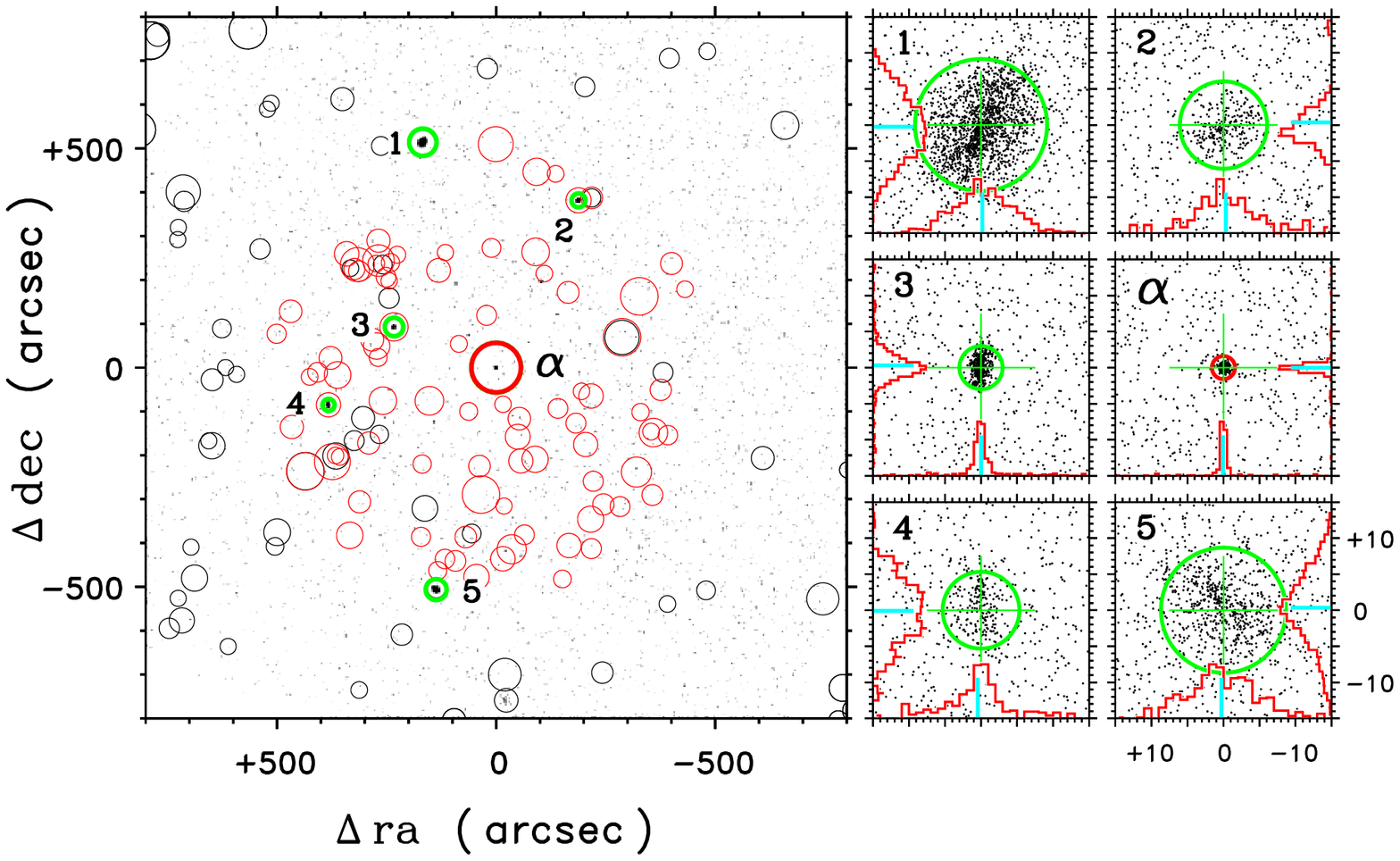} 
\vskip  0mm
\figcaption[]{\small
{\em Left:}\/ {\em Chandra}\/ HRC-I image (dark spots) of the field around $\alpha$~Per, with superimposed astrometric positions: black for A2, red for 2MASS.  The magnitude cutoffs are $V\leq 13$ and $J\leq 13$, for the respective catalogs.  The green-circled objects were used to refine the astrometry of the HRC-I image under the assumption that the X-ray source centroids share the coordinates of the optical counterparts.  Four of the reference sources are Alpha Per members, but the X-ray brightest in the field -- \#{1} (HE\,615) -- is a background object at about twice the cluster distance.  {\em Right:}\/  Astrometric registration procedure.  Each panel depicts the individual X-ray events (small black dots: some are background) of one of the {\em Chandra}\/ sources; the {\em Gaia}\/ optical position by the green circle (95\% encircled energy radius) and cross-hairs; and the (refined) X-ray position by blue tick marks (indicating the centroid of the events profile [red histogram] in sky $\Delta{\rm ra}$ or $\Delta{\rm dec}$).  The event clouds become progressively more blurred the further off-axis the sources are.  The central, well-focused region around $\alpha$~Per shows a single, sharp point source, coincident with the coordinates of the bright star; confirming -- what already had been anticipated by the optical/NUV imaging -- that the {\em ROSAT}\/ offset was a false positive.
}
\end{figure}

\subsubsection{{\em Chandra}\/ HRC-I Imaging}

To resolve the high-energy environs around $\alpha$~Per, {\em Chandra}\/ was the X-ray telescope of choice, with its excellent 1$\arcsec$ imaging (compared to, say, the $\sim$10$\arcsec$ resolution of {\em XMM-Newton}\,).  {\em Chandra}\,'s High Resolution Camera (HRC-I) was the best option of the two available imagers  on the observatory, with its large $30\arcmin{\times}30\arcmin$ field of view and excellent low-energy response (important for stellar coronal sources).  The other camera system, CCD-based ACIS, has poorer soft response and a ``red leak'' that hampers its use for optically bright objects like $\alpha$~Per.  The low -- essentially nonexistent -- energy resolution of HRC-I was not a concern, because {\em ROSAT}\/ already had shown that the main source is hard: if $\alpha$~Per turned out to be a faint secondary source, there likely would not be enough counts in a reasonable-length observation to perform a spectral analysis.

The HRC-I pointing on $\alpha$~Per was carried out 2015 December 22, about a month after the first {\em HST}\/ WFC3 visit.  The approved exposure was 20~kiloseconds (ks), with about 21~ks collected during the actual observation.  The source detected by {\em ROSAT}\/ was expected to yield at least 300 counts, and this would provide a sub-arcsecond X-ray position.  For any other source close to the field center, the accumulated cosmic background in a 20~ks exposure would be low (a few counts in a $r=1.6\arcsec$ detect cell [95\% encircled energy]), so 10 net counts would be a secure detection.  This corresponds to a $L_{\rm X}$ threshold of $\sim1.5\times10^{28}$ ergs s$^{-1}$ (at the cluster center distance, 172~pc), several times deeper than the Prosser et al.\ (1996) Alpha Per Survey with {\em ROSAT},\/ thanks mainly to the higher spatial resolution, and thus better background rejection, of {\em Chandra}. 

The HRC-I image of $\alpha$~Per is illustrated in Figure~5.  The left side of the diagram is a time-integrated X-ray event map (dark spots), binned in 0.5$\arcsec$ spatial pixels, with the $\alpha$~Per source at center (near the observation aimpoint).  Around the central region, several other bright X-ray sources (numbered 1--5) are coincident with optical counterparts from USNO-A2 (black circles) or 2MASS (red circles).  The $V$ magnitudes range from 11.1 for the brightest (\#1: HE\,615, probably early-K [and spectroscopic binary]) to about 14.7 for the faintest, and reddest (\#4: APX~132, probably a late-K dwarf).  All of these are late-type coronal sources, given the optical and 2MASS colors of the counterparts, and all were identified in the earlier {\em ROSAT}\/ imaging by Prosser \& Randich (1998) or Prosser et al.\ (1998).  The properties of the reference objects are summarized in Table~2.

The optical coordinates of the counterparts were exploited to refine the astrometry of the HRC high-energy map.  For this purpose, positions were taken from the first data release of the {\em Gaia}\/ mission\footnote{see: http://www.cosmos.esa.int/web/gaia/dr1}, rather than 2MASS or USNO~A2.  Unlike 2MASS, the {\em Gaia}\/ coordinates have essentially negligible errors; and unlike A2, the epoch of {\em Gaia}\/ is very recent (2015.0), essentially the same as the {\em Chandra}\/ pointing (2015.98), so the possible uncertainty due to proper motion corrections is removed\footnote{Mermilliod et al.\ 2008 derived proper motions for the Alpha Per cluster of $\mu_{\alpha}\cos{\delta}= +0.022\arcsec$~y$^{-1}$ and $\mu_{\delta}= -0.025\arcsec$~y$^{-1}$, so the 1 year difference between the {\em Gaia}\/ epoch and that of the HRC exposure represents a negligible coordinate shift at the level of the typical X-ray centroid uncertainty ($\sim{\pm}0.2\arcsec$: see Table~2).}.  The {\em Gaia}\/ first data release did not include an entry for the bright star $\alpha$~Per, so the {\em Hipparcos}\/ coordinates for epoch 2015 were used instead.  This position should be fully consistent with the {\em Gaia}\/ reference frame.  

The overall procedure is portrayed schematically in the six panels on the right hand side of Fig.~5.  These depict localized regions around each X-ray source, with dots representing individual counts.  Note the blurring of the event clouds of the sources furthest from the aimpoint (at $\alpha$~Per), a characteristic of the curved focal plane of the X-ray telescope recorded by the flat HRC camera.  Each of the sources, flanking central $\alpha$~Per, were centroided in sky coordinates ($\Delta{x}$ [``ra''], $\Delta{y}$ [``dec'']; N is up, E to the left; both axes in arcseconds), whose origin was the optical position of $\alpha$~Per.  

The registration exercise resulted in an absolute shift of the current-epoch {\em Chandra}\/ image by $+0.8\arcsec$ in Right Ascension, and $+0.4\arcsec$ in Declination (corrected in Fig.~5).  These (small) shifts are consistent with the excellent absolute aspect reconstruction typical of {\em Chandra.}\/ The technical {\em relative}\/ offset of the $\alpha$~Per X-ray source from the optical coordinates of the bright star, with respect to the five reference objects, was essentially zero with a 1\,$\sigma$ dispersion (among the five calibrators) of ${\pm}0.2\arcsec$ (in sky $x$ and $y$, independently), or a 1 standard error of the mean (s.e.) of about ${\pm}0.1\arcsec$ (Table~2).  In short, the $\alpha$~Per X-ray source is isolated, single, and coincident with the bright star.  If one now wishes to invoke an active coronal companion to $\alpha$~Per, still a viable option as described below, it would have to be very close to the supergiant, within about a dozen AU.

The $\alpha$~Per source had a total of 392 counts in a $r=1.6\arcsec$ detect cell (95\% encircled energy near the aimpoint) in the 21~ks exposure.  The projected background was 4 counts in that cell (measured from broad areas away from the point sources).  The net count rate is consistent with the {\em ROSAT}\/ observation two decades earlier.  An examination of the events time series for $\alpha$~Per found some evidence of short time scale variability, but no obvious flares.  The X-ray properties of $\alpha$~Per, and the reference 2MASS sources (including their optical parameters from previous work), are summarized in Table~2.  For the off-axis sources, the 95\% encircled energy radius was assumed to scale as $(1.6 + 0.06\,\rho^{2.2})$ ($\arcsec$), where $\rho$ ($\arcmin$) is the displacement of the source from the image center.  

An Energy Conversion Factor of ${\rm ECF}=8.7^{+0.8}_{-0.3}{\times}10^{-12}$ erg cm$^{-2}$ count$^{-1}$ was applied to these hard sources, for count rates adjusted for the encircled energy fraction, to translate the apparent HRC-I count rates into energy fluxes at Earth.\footnote{The upper/lower limits refer to the average of deviations between $\Delta{T}=0.2$~dex and $\Delta{N_{\rm H}}=0.2$~dex, where the central value was calculated for $T= 10^{7}$~K and $N_{\rm H}=2{\times}10^{20}$ cm$^{-2}$, appropriate for $\alpha$~Per with $E(B-V)\sim +0.04$, using the {\em Chandra}\/ WebPIMMS tool (Cycle~17 version) (See: http://cxc.harvard.edu/toolkit/pimms.jsp) for a solar abundance APEC Plasma model and energy range 0.2--2~keV for the unabsorbed flux.}  Specifically for the $\alpha$~Per source, $f_{\rm X}\sim 1.7{\times}10^{-13}$ erg cm$^{-2}$ s$^{-1}$ (0.2--2~keV), or equivalently $L_{\rm X}\sim 5\times10^{29}$ erg s$^{-1}$ at the assumed 155~pc distance.

The X-ray luminosities listed in Table~2 are systematically lower than reported in the previous Alpha Per studies by Prosser and collaborators, by about 0.4~dex.  However, the HRC-I count rates (CR) here are nearly identical to the PSPC CR measured nearly a quarter century ago, for all the targets except the non-member HE\,615, which has about twice the CR in the present epoch.  According to simulations with WebPIMMS, the conversion between PSPC 0.1--2~keV counts (utilized by Prosser et al.) and HRC-I counts is essentially unity with only a slight dependence on $N_{\rm H}$ (between the $5{\times}10^{20}$ cm$^{-2}$ of the previous work and the $2{\times}10^{20}$ cm$^{-2}$ assumed here) for hard sources with $T\sim 10$~MK.  In fact, the systematically higher $L_{\rm X}$ of the previous Alpha Per survey can be traced partly to the broader reference energy band (0.1--2~keV versus 0.2--2~keV here), but mainly to the two-temperature (0.2~keV\,+\,1.0~keV) ECF utilized for the {\em ROSAT}\/ PSPC at the time (${\rm ECF}= 2{\times}10^{-11}$ erg cm$^{-2}$ count$^{-1}$: Prosser et al.\ 1996), compared with a 1--$T$ version equivalent to that adopted here (${\rm ECF}\sim 1{\times}10^{-11}$ erg cm$^{-2}$ count$^{-1}$ for $T=1\times10^{7}$~K and $N_{\rm H}= 2{\times}10^{20}$ cm$^{-2}$).  

The slightly softer energy band adopted for the PSPC in the Prosser et al.\ studies, together with the strong influence of the higher assumed ISM absorption on the lower$-T$ component, boosted the overall ECF by almost a factor of two compared with the $1-T$ lower-absorption counterpart.  The higher low-energy cutoff adopted for the HRC-I here (0.2~keV) makes it somewhat less sensitive to soft absorption, and so the ECF for a $2-T$ model equivalent to that of Prosser et al.\ is only about 10\% higher than the $1-T$ version (at least for $N_{\rm H}= 2{\times}10^{20}$ cm$^{-2}$).  These differences highlight the large systematics in $L_{\rm X}$ that can result from different assumptions concerning the underlying ECFs, as well as the reference energy bands.  However, the {\em relative}\/ behavior among a sample of objects (as described later) should be maintained, if a consistent set of assumptions describing the ECFs of the coronal sources is adopted (see Appendix B).  To be sure, factors of two systematics pale in comparison to the many orders of magnitude spanned by the coronal luminosities of quiet and active stars in general.

\begin{deluxetable}{clrcrrrcrccc}
\rotate
\tabletypesize{\footnotesize}
\tablenum{2}
\tablecaption{{\em Chandra}\/ HRC-I Sources in $\alpha$~Per Field}
\tablecolumns{12}
\tablewidth{0pt}
\tablehead{\colhead{Target No.} & \colhead{Name} & \colhead{2MASS} & \colhead{$\rho$} & \colhead{$\Delta{x}$} & 
\colhead{$\Delta{y}$} & \colhead{$V$} & 
\colhead{$(B-V)$} & \colhead{$J$}   & \colhead{Sp.~Typ.}  &  \colhead{Count Rate} &  \colhead{$\log{L_{\rm X}}$} \\
\colhead{} & \colhead{} & \colhead{} & \colhead{$(\arcmin)$}& \colhead{$(\arcsec)$} & 
\colhead{$(\arcsec)$} & \colhead{(mag)} & 
\colhead{(mag)} & \colhead{(mag)}   & \colhead{}  &  \colhead{(counts ks$^{-1}$)} & \colhead{(erg s$^{-1}$)} \\
\colhead{(1)} & \colhead{(2)} & \colhead{(3)} & \colhead{(4)} & \colhead{(5)}  & \colhead{(6)}  & \colhead{(7)} & \colhead{(8)} & \colhead{(9)}  & \colhead{(10)}  & \colhead{(11)} & \colhead{(12)} 
} 
\startdata
$\alpha$ & $\alpha$~Per                  & J03241936+4951401 & 0.0 & $+0.8$ & $+0.4$ & +1.79  & +0.48  &  +0.83 & F5~Ib & $19.5{\pm}1.0$ & 29.69 \\
       1 & HE\,615\tablenotemark{a}     & J03243674+5000139 & 9.0 & $+0.7$ & $+0.2$ & +11.06  & +0.90  &  +9.24 & K0~IV?  & $70.9{\pm}1.9$ & 30.86 \\
       2 & APX\,120A                               & J03235988+4958017 & 7.1 & $+0.4$ & $+0.1$ & +13.94  & +1.30  & +11.29 & K6~V  & $11.0{\pm}0.7$ & 29.53 \\
       3 & APX\,27A                                & J03244348+4953130 & 4.2 & $+0.9$ & $+0.4$ & +12.58  & +0.92  & +10.74 & K1~V  & $20.6{\pm}1.0$ & 29.80 \\
       4 & APX\,132A                               & J03245892+4950152 & 6.5 & $+1.0$ & $+0.6$ & +14.69  & +1.35  & +11.47 & K7~V  &  $8.7{\pm}0.7$ & 29.43 \\  
       5 & APX\,24                                 & J03243348+4943138 & 8.7 & $+0.8$ & $+0.7$ & +12.07  & +0.82  & +10.39 & G9~V  & $24.9{\pm}1.1$ & 29.88 \\ 
\hline
          \multicolumn{4}{r}{Event List Average Shifts (excl.\ $\alpha$~Per):} &  $+0.8$ & $+0.4$ \\ 
          \multicolumn{4}{r}{(${\pm}1\,\sigma$):} &  ${\pm}0.2$ & ${\pm}0.2$ \\ 
          \multicolumn{4}{r}{$\alpha$~Per Relative Shifts:} &  $+0.0$ & $+0.0$ \\          
          \multicolumn{4}{r}{(${\pm}1\,{\rm s.e.}$):} &  ${\pm}0.1$ & ${\pm}0.1$ \\ 
\enddata
\tablenotetext{a}{Non-member; SB1, probably RS~CVn binary given the high $L_{\rm X}$.}
\tablecomments{Parameters for $\alpha$~Per from SIMBAD; all others from Prosser \& Randich (1998) or Prosser et al.\ (1998); except Col.~9 near-IR magnitudes from 2MASS, and Col.~10 spectral types, which were estimated from $(B-V)_0$ (with $E(B-V)\sim +0.04$) according to Fitzgerald (1970).  Col.~4: $\rho$ is the approximate displacement of the source from the image center.  Col.~11 net count rates were corrected for the 95\% encircled energy factor.  Col.~12 X-ray luminosities based on $d=155$~pc for $\alpha$~Per, but $d=172$~pc (cluster center) for the other objects (excluding non-member HE\,615, whose {\em Gaia}\/ distance is 313~pc).  The 1\,$\sigma$ uncertainty on $\Delta\rho\equiv \sqrt{\Delta{x}^2\,+\,\Delta{y}^2}$ is approximately FWHM/(S/N) (Ayres 2004), where FWHM is the diameter of the point spread function at the half-intensity points.  For the HRC-I pointing on $\alpha$~Per, $\sigma_{\Delta\rho}$ is roughly $(1\,+\,0.04\,\rho^{2.2})/\sqrt{21{\times}{\rm CR}}$ ($\arcsec$), where CR is in counts ks$^{-1}$.  For $\alpha$~Per, with the minimum FWHM, $\sigma_{\Delta\rho}\sim 0.05\arcsec$; while for the other sources it ranges from 0.1$\arcsec$--0.3$\arcsec$.  If $\sigma_{\Delta{x}}$ and $\sigma_{\Delta{y}}$ are uncorrelated, then individually they should be about 70\% of $\sigma_{\Delta\rho}$, similar to the standard deviations of the $\Delta{x}$ and $\Delta{y}$ values over the five calibrators (${\pm}0.2\arcsec$).  At this level, the uncertainties in the {\em Gaia}\/ coordinates (epoch~2015) are negligible.
}
\end{deluxetable}

\clearpage
\section{ANALYSIS}

The left hand side of Figure~6 compares $L_{\rm X}/L_{\rm bol}$ versus $L_{\rm Si\,IV}/L_{\rm bol}$ for selected G, K, and M dwarfs and F and G supergiants.  The heritages of the flux measurements are summarized in Appendix B.  The X-ray values listed in Table~B2 were corrected for reddening as a natural consequence of the CR to flux conversion process.  The FUV values, on the other hand, were measured directly from the observed flux-calibrated spectra, then corrected for reddening post facto, according to the wavelength-dependent average galactic extinction formula of Fitzpatrick \& Massa (2007).

As noted earlier, the bolometric normalization potentially allows a fairer comparison among stars of different sizes and distances.  Spectral types and luminosity classes are encoded in the figure by symbols and colors according to the two legends.  The main subject of the study, $\alpha$~Per, and comparison F supergiant Canopus are the black-bordered diamonds at the left hand side of the panel ($\alpha$~Per is the larger one).  The symbols outlined in red are Cepheid variables.  ``Sideways L'' shaded bars connect the Cepheid FUV low and high states for the average X-ray low state, and the X-ray low and high states for the average FUV low state (see Appendix B: the brief X-ray high states seen in two Cepheids apparently are out of phase with the FUV enhancements, which occur near the optical ``piston'' phase [Engle et al.\ 2014; Ruby et al.\ 2016; Engle 2015]).  The gray shaded wedge is meant to depict the power-law relation between the coronal and subcoronal diagnostics followed loosely by G dwarfs, while the elongated blue hatched oval highlights the approximate relationship obeyed by G supergiants.  The G supergiants are ``X-ray deficient'' with respect to the G dwarfs (or, ``\ion{Si}{4} super-luminous,'' depending on how one views the displaced power laws).  Note that the later dwarfs follow essentially the same power law slope as the G dwarfs, although perhaps shifted slightly to the left for the coolest objects.  Note, also, that both $\alpha$~Per and Canopus sit far to the left of the apparent G supergiant power law, although perhaps not so far from the region defined by the two Cepheid X-ray high states (at the FUV low states).  

\clearpage
\begin{figure}[ht]
\figurenum{6}
\vskip  0mm
\hskip -5mm
\includegraphics[scale=0.75,angle=90]{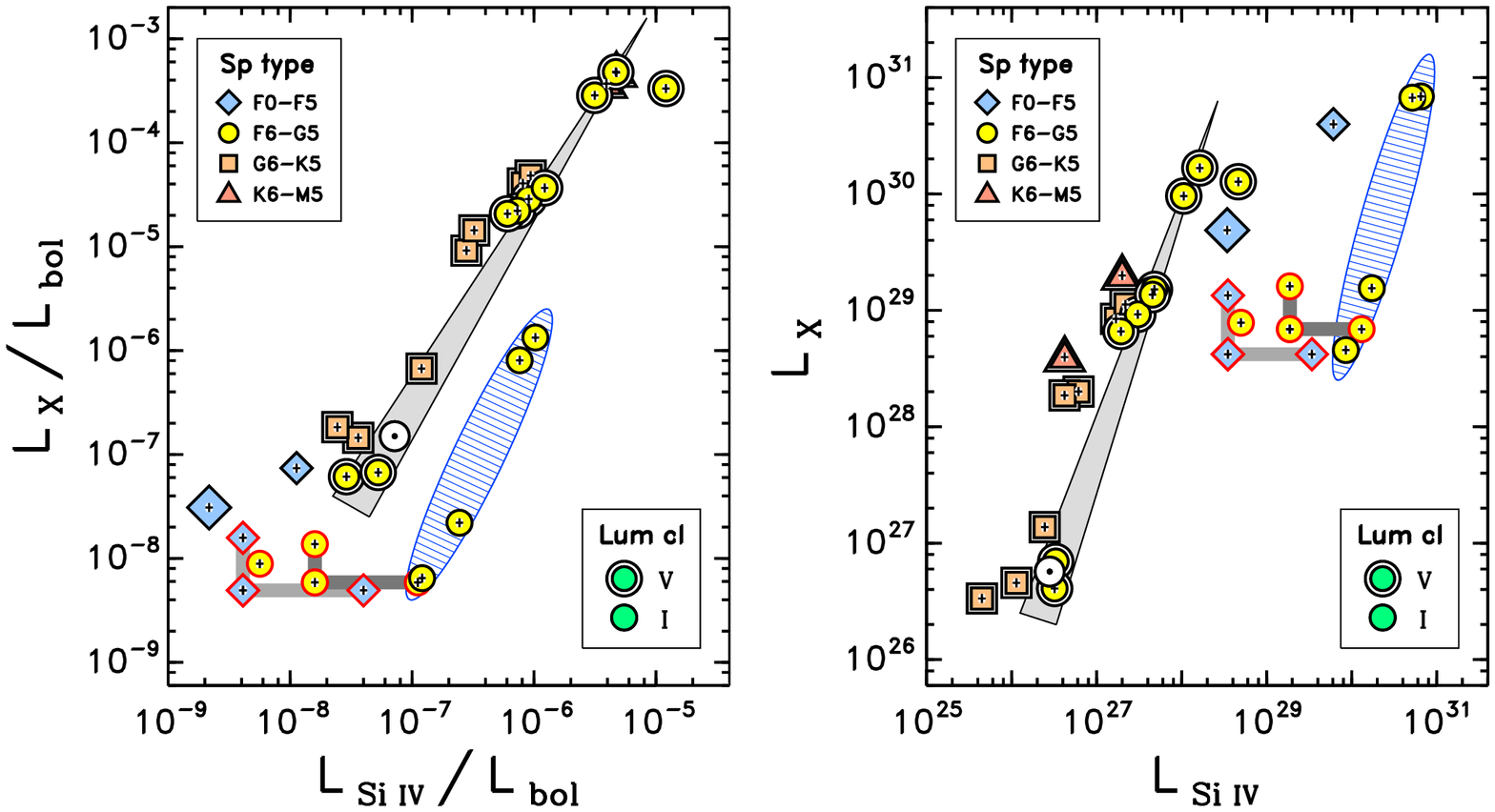} 
\vskip  0mm
\figcaption[]{\small
{\em Left:}\/ $L_{\rm X}/L_{\rm bol}$ versus $L_{\rm Si\,IV}/L_{\rm bol}$ for G--M dwarfs and F and G supergiants, according to the two legends.  {\em Right:}\/  Alternative X-ray/{Si}\,{\scriptsize IV} diagram, expressed in absolute luminosities.  In both panels, the Sun is marked $\odot$; $\alpha$~Per is the larger of the two isolated, black bordered blue diamonds; and Canopus is the smaller of the two.
}
\end{figure}

The right hand side of Fig.~6 is an alternative version of the coronal/subcoronal diagram, now expressed in absolute luminosities.  The G dwarfs still appear to follow a coherent power law, as do the G supergiants, and the conspicuous displacement between the two remains.  Now, however, the K and M dwarfs display perhaps more of a leftward separation with respect to the G dwarf power law.  The Cepheid FUV high states (at the X-ray low states) still connect to the lower end of the G supergiant power law.  However, the Cepheid FUV low states (together with the X-ray high states) now fall between the G dwarf and G supergiant relations.  The two non-variable F supergiants also lie in the same gap, although higher up in $L_{\rm X}$ ($\alpha$~Per is the larger diamond).  At the same time, it is clear that the X-ray luminosity of $\alpha$~Per is similar to that of a very active G dwarf, but perhaps higher than the most active K and M dwarfs; while the $L_{\rm X}$ of Canopus would correspond to the very most active G dwarfs (none of which are represented in this nearby stellar sample, but have been observed, at least in X-rays, in the very young [$\sim 13$~Myr] h~Persei cluster [Argiroffi et al.\ 2016]).

The X-ray/\ion{Si}{4} diagrams demonstrate that both $\alpha$~Per and Canopus are discrepant with respect to normal G supergiants of all activity levels ($\alpha$~Aqr and $\beta$~Aqr at the low end, $\beta$~Dra and $\beta$~Cam at the high end).  However, the nature of the discrepancy (X-ray over-luminous or \ion{Si}{4} under-luminous) cannot be decided by this comparison alone.

\clearpage
\begin{figure}[ht]
\figurenum{7}
\vskip  0mm
\hskip -5mm
\includegraphics[scale=0.75,angle=90]{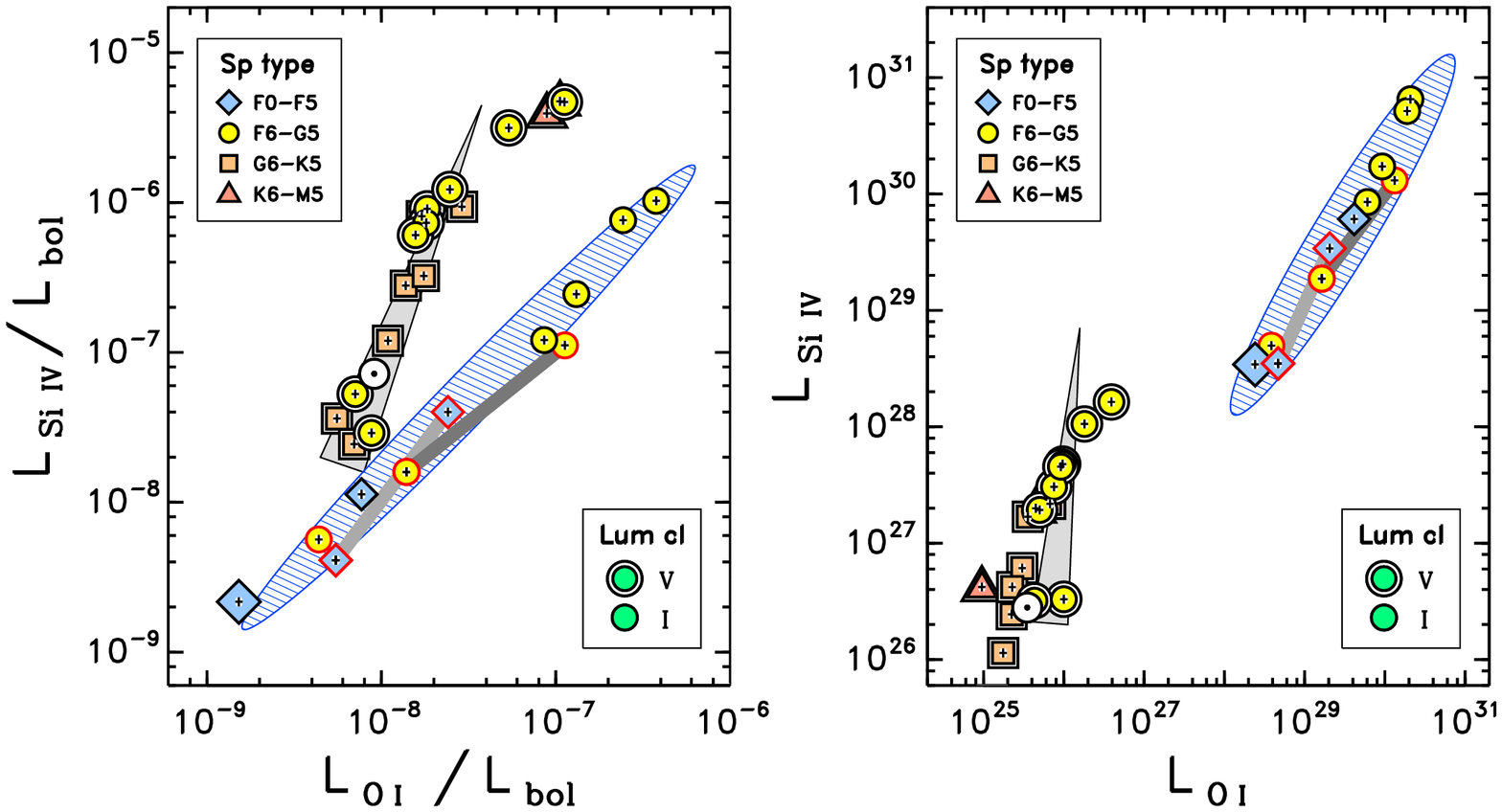} 
\vskip  0mm
\figcaption[]{\small
{\em Left:}\/ $L_{\rm Si\,IV}/L_{\rm bol}$ versus $L_{\rm O\,I}/L_{\rm bol}$ for G--M dwarfs and F and G supergiants, according to the two legends.  {\em Right:}\/ Alternative {Si}\,{\scriptsize IV}/{O}\,{\scriptsize I} diagram, expressed in absolute luminosities.  In both panels, the Sun is marked $\odot$; $\alpha$~Per is the larger of the two isolated, black bordered blue diamonds; and Canopus is the smaller of the two.
}
\end{figure}

Some clarity can be achieved by considering an analogous pair of diagrams, but now for subcoronal \ion{Si}{4} versus chromospheric \ion{O}{1}] 1355~\AA.  The intercombination line is favored over the 1305~\AA\ resonance lines because it likely is optically thin and thus less affected by blueshifted absorption in the windy supergiants like $\alpha$~Aqr.  Further, the 1305~\AA\ multiplet often is contaminated by atomic oxygen skyglow in at least COS spectra.  

The \ion{Si}{4}/\ion{O}{1} comparison is illustrated in Figure~7.  The bolometric normalized fluxes in the left hand panel show some of the same regularities displayed in the previous figure, namely the dwarf stars tend to fall on a coherent power law relation, while the supergiants also follow a power law, but displaced from the dwarf version.  Here, however, the slopes of the two power laws are quite different, much shallower for the luminous stars.  Curiously, the connected FUV low and high states of the Cepheids seem to track the overall G supergiant trend quite well.  Further, the most active G and M dwarfs appear to break away from the main power law, possibly indicating ``saturation'' of subcoronal \ion{Si}{4} (e.g., Vilhu \& Rucinski 1983).  Most importantly, both $\alpha$~Per (larger blue diamond) and Canopus (smaller black-bordered diamond) arguably fall on a lower extension of the G supergiant relation, rather than sitting in an entirely disconnected region of the diagram as in Fig.~6.

The comparison on the right hand side of Fig.~7, involving the absolute luminosities, is perhaps more illuminating.  Here, there is a clear separation between the dwarf-star behavior (low to moderate subcoronal luminosities, low chromospheric oxygen luminosities) and the supergiant counterpart (moderate to high subcoronal luminosities, moderate to high chromospheric luminosities), with minimal overlap in $L_{\rm Si\,IV}$ between the two distributions (seen also in the right hand panel of Fig.~6, but more obvious here).  The strong correlation of the subcoronal and chromospheric emissions in the non-variable G supergiants as well as the Cepheids, and the fact that both $\alpha$~Per and Canopus fall on the same relationship, implies that the \ion{Si}{4} levels of the two F supergiants are fully consistent with their chromospheric emissions.  Thus, the major impetus for the displaced locations of both stars in the coronal/subcoronal diagram (Fig.~6) must come from the X-ray side.

\section{DISCUSSION}

On the surface, the lack of X-ray active visual companions supports the idea that both F supergiants are members of a novel class of coronal emitters, characterized by low $L_{\rm Si\,IV}/L_{\rm X}$ ratios compared to normal G-type supergiants.  However, the case against active dwarf companions, especially for $\alpha$~Per, is not completely closed.  The X-ray imaging, in particular, does not exclude a close (``spectroscopic'') companion, which given the great distances of the supergiants still could be dozens of AU from the primary, impossible to detect interferometrically with current instrumentation (given the $\Delta{V}\sim 9$ contrast), and a challenge even for state-of-the-art Doppler-reflex spectroscopy (especially given the high level of radial velocity jitter in such stars [e.g., Hatzes \& Cochran 1995; Lee et al.\ 2012]).  In fact, the X-ray luminosity of $\alpha$~Per is similar to G dwarfs of the cluster, and the extrapolated \ion{Si}{4} luminosity of a putative companion of that type would be {\em less}\/ than what appears to be the intrinsic \ion{Si}{4} emission of $\alpha$~Per, so easily could be hidden.  The case for a close companion to Canopus is more tenuous (owing to the higher $L_{\rm X}$), but still tenable. 

\section{FOR THE FUTURE}

A possible path forward to resolve the apparent coronal conundrum would be to collect additional examples of X-ray and FUV emissions of early-F supergiants straddling the range bounded by Canopus on the warm side and $\alpha$~Per on the cool side.  Although F supergiants are relatively rare (massive stars in a transient phase of evolution), there nevertheless are a number of potential candidates similar in brightness to $\alpha$~Per and with low reddening: both aspects are vital for successful FUV measurements.  If the same X-ray anomalies seen in $\alpha$~Per and Canopus appear in additional examples, confidence will grow that a novel mechanism of coronal excitation must be considered.  For Canopus, the apparent $L_{\rm X}$ is close to, and perhaps exceeds, the upper bound seen in very young ``X-ray saturated'' late-type G dwarfs.  If an early-F supergiant can be found with a significantly higher $L_{\rm X}$, then the case in favor of a new coronal paradigm at the edge of convection would be solidified.

\acknowledgments
This work was supported by grant GO6-17005X from the Smithsonian Astrophysical Observatory, based on observations from the {\em Chandra}\/ X-ray Observatory, collected and processed at the {\em Chandra}\/ X-ray Center, operated by SAO under contract to NASA; and grants GO-14349 ($\alpha$~Per) and GO-12278 (Advanced Spectral Library [ASTRAL]: Cool Stars) from the Space Telescope Science Institute, based on observations from {\em Hubble Space Telescope}\/ collected at STScI, operated by the Associated Universities for Research in Astronomy, also under contract to NASA.  Ground-based observations were obtained with the Apache Point Observatory 3.5~m telescope, operated by the Astrophysical Research Corporation.  Dr.\ Carol Grady guided the APO/GIFS imaging, and generously contributed time for the project out of her engineering allocation.  This study also accessed public databases hosted by {SIMBAD}, maintained by {CDS}, Strasbourg, France; the Mikulski Archive for Space Telescopes at STScI in Baltimore, Maryland; and the High Energy Astrophysics Science and Research Center at the NASA Goddard Space Flight Center, in Greenbelt, Maryland.  Additionally, this work has made use of data from the European Space Agency (ESA) mission {\it Gaia} (\url{http://www.cosmos.esa.int/gaia}), processed by the {\it Gaia} Data Processing and Analysis Consortium (DPAC, \url{http://www.cosmos.esa.int/web/gaia/dpac/consortium}). Funding
for the DPAC has been provided by national institutions, in particular the institutions participating in the {\it Gaia} Multilateral Agreement.     

\clearpage
\appendix

\section{Stellar Parameters}

Although not initially a main focus of the $\alpha$~Per multi-wavelength campaign, the flux-flux comparisons illustrated in Figs.~6 and 7 provided a helpful context for evaluating the unusual $L_{\rm Si\,IV}/L_{\rm X}$ ratios of the cluster supergiant.  The comparison stars, and their properties, are described in this Appendix.

The sample consists of two separate classes of objects: (1) G, K, and M dwarfs, chosen because of the potential role of a Main sequence companion in contributing to the X-rays of $\alpha$~Per; and (2) F and G supergiants, the most relevant class for direct comparisons to $\alpha$~Per itself (and related object Canopus).  The key parameter for the flux-flux comparisons are bolometric fluxes, $f_{\rm bol}$ (erg cm$^{-2}$ s$^{-1}$ at Earth), which are utilized as normalizing factors for the measured X-ray and FUV intensities.  The bolometric fluxes were calculated as,
\begin{equation}
f_{\rm bol}= 2.53{\times}10^{-5}\,{\times}\,10^{-(V\,+\,B.C.)/2.5}\,\,,
\end{equation}
based on the solar parameters of Bessell et al.\ (1998) and the solar luminosity cited by Ayres et al.\ (2006).  Here, $V$ is the de-reddened visual magnitude and B.C.\ is the bolometric correction.  The leading coefficient in the relation is slightly lower than proposed by Ayres et al.\ (2005), reflecting a slight downward revision in the average solar absolute irradiance (see discussion in Ayres et al.\ 2006).  From the $f_{\rm bol}$ and distances (say, from {\em Hipparcos}\/ parallaxes), one can deduce absolute luminosities, $L_{\rm bol}$.  Then, given effective temperatures, $T_{\rm eff}$, derived from colors or other considerations (e.g., model atmospheres and spectral diagnostics: see, e.g., PASTEL catalog of Soubiran et al.\ [2016]), one can construct an H--R diagram (for example, Fig.~1, here). 

Table~A1 lists the sample stars and their properties, as derived mainly from SIMBAD, but also from other sources as noted.  The stars were selected to have modern FUV fluxes (i.e., from {\em HST}\,), and good quality X-ray measurements from {\em ROSAT,}\/ {\em Chandra,}\/ and/or {\em XMM-Newton.}  For the purposes here, the sample does not need to be complete, but rather representative.  In terms of the fundamental parameters, directly measured visual magnitudes and colors are straightforward to find, but derivative color excesses, say $E(B-V)$, and bolometric corrections are more challenging, and conflicting values often appear in the literature.  Again, for the purposes here, it is not essential to argue whether a selected parameter from a distribution of values is the most correct, but rather that statistical deviations of the chosen parameters are small enough that trends illuminated by the sample would be preserved (which empirically appears to be the case in Figs.~6 and 7, helped, of course, by the many-decade logarithmic axes scales).

Color excesses were collected mainly for the supergiants, because, aside from two more distant cluster dwarfs, all the other Main sequence stars are close enough that reddening should be negligible.  To derive de-reddened visual magnitudes, a Galactic reddening law $A_{V}= 3.1\,E(B-V)$ was assumed.  Bolometric corrections were estimated from re-reddened $B-V$ colors, according to the transformations proposed by Flower (1996).  These transformations begin to fall apart among the later spectral types, for which infrared colors are more appropriate than the shorter-wavelength bands for deriving bolometric corrections.  However, for the two latest objects (dMe flare stars), the bolometric corrections were calculated such that the $f_{\rm bol}$ relation would predict the correct $L_{\rm bol}$, known from other considerations.

\begin{deluxetable}{lclccccccc}
\rotate
\tabletypesize{\small}
\tablenum{A1}
\tablecaption{Comparison Stars and Stellar Parameters}
\tablecolumns{10}
\tablewidth{0pt}
\tablehead{\colhead{Name} & \colhead{HD~No.} & \colhead{Type} & 
\colhead{$V$}  & \colhead{$(B-V)$}  &  \colhead{$E(B-V)$} &  \colhead{B.C.} &  \colhead{$T_{\rm eff}$} &  
\colhead{$d$}  &  \colhead{Notes}  \\
\colhead{} & \colhead{} & \colhead{} & \colhead{(mag)} &  \colhead{(mag)} &  \colhead{(mag)} &  \colhead{(mag)} & 
\colhead{(K)} & \colhead{(pc)} & \colhead{}  \\
\colhead{(1)} & \colhead{(2)} & \colhead{(3)} & \colhead{(4)} & \colhead{(5)}  & \colhead{(6)}  & \colhead{(7)} 
 & \colhead{(8)}   & \colhead{(9)}   & \colhead{(10)}  
} 
\startdata
 $\alpha$~Car                         &   45348 & {\em F0\,Ib} & $-0.74$  & $+0.15$ & $+0.01$ & $+0.03$ & 7410 & 94.8 & 1,2 \\ 
  \\
 $\alpha$~Per                         &   20902 & F5\,Ib & $+1.79$  & $+0.48$ & $+0.04$ & $-0.01$ & 6350 & 155  & 1,2\\ 
  \\
 $\delta$~Cep\tablenotemark{a}        &  213306 & F5\,Iab & $+3.75$  & $+0.60$ & $+0.07$ & $-0.04$ & {\em 5840} & 265  & 3,4,5,6,7  \\ 
  \\
 $\alpha$~UMi\tablenotemark{a}        &    8890 & F8\,Ib & $+2.02$  & $+0.60$ & $+0.00$ & $-0.06$ & 5930 & 133  & 2,4,8  \\ 
  \\
 $\beta$~Dor\tablenotemark{a}         &   37350 & F8/G0\,Ib & $+3.76$  & $+0.82$ & $+0.04$ & $-0.15$ & 5440 & 310  & 1,8 \\ 
  \\
 $\beta$~Aqr                          &  204867 & G0\,Ib & $+2.89$  & $+0.82$ & $+0.02$ & $-0.16$ & 5480 & 165   & 1,2 \\ 
  \\
 $\beta$~Cam                          &   31910 & G1\,Ib--II & $+4.02$  & $+0.93$ & $+0.10$  & $-0.18$ & 5440 & 270   & 9,10 \\ 
  \\
 $\beta$~Dra                          &  159181 & G2\,Ib--IIa & $+2.81$  & $+0.98$ & $+0.10$ & $-0.21$ & 5290 & 117   & 1,2,10 \\ 
  \\
 $\alpha$~Aqr                         &  209750 & G2\,Ib & $+2.94$  & $+0.96$ & $+0.02$ & $-0.26$ & 5250 & 161   & 1,2 \\ 
  \\
 $\zeta$~Dor                          &   33262 & F9\,V  & $+4.72$  & $+0.47$ & \nodata  & $-0.00$ & 6160 & 11.6   &  \\ 
  \\
 $\chi$~Her                           &  142373 & G0\,V  & $+4.62$  & $+0.57$ & \nodata  & $-0.05$ & 5860 & 15.9   & \\ 
  \\
 $\chi^1$~Ori                         &   39587 & G0\,V  & $+4.40$  & $+0.60$ & \nodata  & $-0.06$ & 5950 & 8.66    &  \\ 
  \\
 EK~Dra                               &  129333 & G1.5\,V  & $+7.61$  & $+0.59$ & \nodata  & $-0.06$ & 5750 & 34.1   &  \\ 
  \\
 $\pi^1$~UMa                          &   72905 & G1.5\,V  & $+5.64$  & $+0.62$ & \nodata  & $-0.07$ & 5880 & 14.4   &  \\ 
  \\
 H\,II\,314                 & (V1038 Tau) & G1--2\,V  & $+10.66$ & $+0.66$ & $+0.04$ & $-0.07$ & {\em 5600} & {\em 133}  &  Pleiades; 11  \\ 
  \\
 Sun                                  &   {\nodata} & G2\,V  & {\em --26.76}  & $(+0.62)$ & \nodata  & $-0.07$ & {\em 5772} & 1~AU &  12,13 \\ 
  \\
 HE\,699                   & (V532 Per) & G2-3\,V  & $+11.27$ & $+0.71$ & $+0.09$ & $-0.07$ & {\em 5550} & {\em 172}   &  Alpha Per; 11 \\ 
  \\
 $\alpha$~Cen~A                       &  128620 & G2\,V  & $+0.01$  & $+0.71$ & \nodata  & $-0.13$ & 5790 & {\em 1.34}  &  14 \\ 
  \\
 $\kappa^1$~Cet                       &   20630 & G5\,V  & $+4.85$  & $+0.67$ & \nodata  & $-0.10$ & 5710 & 9.14  & \\ 
  \\
 61~Vir                               &  115617 & G7\,V  & $+4.74$  & $+0.70$ & \nodata  & $-0.12$ & 5550 & 8.56  &  \\ 
  \\
 $\xi$~Boo~A                          &  131156A & G7\,V  & $+4.68$  & $+0.72$ & \nodata  & $-0.14$ & 5480 & 6.71  &   \\ 
  \\
 $\tau$~Cet                           &   10700 & G8.5\,V  & $+3.50$  & $+0.72$ & \nodata  & $-0.14$ & 5330 & 3.65  &  \\ 
  \\
 HR\,8                                &     166 & K0\,V  & $+6.13$  & $+0.75$ & \nodata  & $-0.16$ & 5460 & 13.7  &  \\ 
  \\
 70~Oph~A                             &  165341A & K0\,V  & $+4.12$  & $+0.85$ & \nodata  & $-0.24$ & 5260 & 5.08  &  \\ 
  \\
 $\alpha$~Cen~B                       &  128621 & K1\,V  & $+1.33$  & $+0.88$ & \nodata  & $-0.27$ & 5230 & {\em 1.34}  & 14  \\ 
  \\
 $\epsilon$~Eri                       &   22049 & K2\,V  & $+3.73$  & $+0.88$ & \nodata  & $-0.27$ & 5080 & 3.22  &   \\ 
  \\
 AU~Mic                               &  197481 & M1\,V  & $+8.62$  & $+1.42$ & \nodata  & {\em --1.50}\,\tablenotemark{b}  & {\em 3630} & 9.9 & 15 \\ 
  \\
 AD~Leo                               & \nodata & M4\,V  & $+9.43$  & $+1.54$ & \nodata  & {\em --2.44}\,\tablenotemark{c}  & {\em 3470} & 4.7 & 15 \\ 
  \\
\enddata
\tablenotetext{a}{Classical Cepheid.}
\tablenotetext{b}{Based on $L_{\star}\sim 0.11 L_{\odot}$ from $T_{\rm eff}$ and $R_{\star}= 0.83\,R_{\odot}$\,$^{(16)}$.}
\tablenotetext{c}{Based on $L_{\star}\sim 0.028 L_{\odot}$\,$^{(15)}$. }
\tablecomments{Col.~3 Types, Col.~4 magnitudes, Col.~5 colors, and Col.~9 distances were extracted from SIMBAD, except as noted (italicized values, references in Notes).  Col.~6 color excesses are from a variety of sources, as noted in Col.~10.  Col.~7 bolometric corrections are from the B.C.--$(B-V)$ transformations of Flower (1996), except as noted.  Col.~8 effective temperatures are from the PASTEL catalog (Soubiran et al.\ 2016), median (many entries) or geometric means (few entries), except as noted.\\[3mm]
{\em References:}\/ (1) Bersier (1996); (2) Kovtyukh et al.\ (2008); (3) Laney \& Caldwell (2007); (4) Bersier (2002); (5) Kovtyukh et al.\ (2010); (6) Kiss \& Szatmary (1998); (7) Krockenberger et al.\ (1998); (8) Kervella et al.\ (2004); (9) Fernie (1982); (10) Luck (2014); (11) Wright et al.\ (2011); (12) Bessell et al.\ (1998); (13) Ayres et al.\ (2006); (14) S{\"o}derhjelm (1999); (15) Maldonado et al.\ (2015); (16) White et al.\ (2015).}
\end{deluxetable}

\section{X-ray and Far-Ultraviolet Fluxes}

Diagnostic fluxes of the comparison stars were derived from archival material, exclusively from {\em HST}\/ for the FUV range, but from several recent X-ray observatories, as mentioned in Appendix~A.  

\subsection{X-rays}

For the X-ray fluxes, a key consideration is the Energy Conversion Factor (ECF), to transform a measured count rate in an instrumental energy range (the band in which the particular camera is sensitive, e.g., 0.1--2.4~keV for the {\em ROSAT}\/ All-Sky Survey [RASS] scans with the PSPC) into an ``instrument-agnostic'' energy range, say 0.2--2~keV.  Disagreements between $L_{\rm X}$ proposed for a given object using different instruments often can be traced to incompatible reported energy bands, or inappropriate ECFs (possible, for example, when observing a very soft X-ray source with an instrument more sensitive to harder photons: see discussion by Ayres et al.\ [2008]).  Another important uncertainty in X-ray measurements is that stellar coronal sources are inherently variable, displaying occasional flares lasting from minutes to hours, rotational modulations on timescales of days to months, and long-term activity cycles on order of years to decades.  The X-ray cycle amplitudes (minimum to peak) can reach factors of 6--10 for even low-activity sunlike stars (e.g., Ayres 2015b: his Fig.~1).  The incessant variability of coronal sources renders snapshot observations of X-ray stars inherently uncertain, far beyond the small statistical errors associated with a high-count detection.  Having multiple epochs of X-ray measurements is, of course, better; but long-term X-ray campaigns are an observational rarity.  Thus the X-ray fluxes reported here should be considered with the multi-scale variability caveat in mind.  The same concern applies also to the FUV, especially since contemporaneous X-ray/FUV measurements are rare as well.  Nevertheless, the extreme contrast of X-ray behavior between young active objects, and their older lower activity cousins, tends to counteract the inherent variability of the sources, and allows broad trends such as in Figs.~6 and 7 to emerge clearly.

The X-ray fluxes reported here were derived from direct observations by the author, and colleagues, using the {\em Chandra}\/ HRC-I, or from the extensive source catalogs of {\em ROSAT}\/ and {\em XMM-Newton},\/ hosted by the High Energy Astrophysics Science and Research Center (HEASARC).  For {\em ROSAT,} there are several possible types of count rates: PSPC values from the All-Sky Survey (RASS); PSPC values from pointed observations; and HRI\footnote{High Resolution Imager, a secondary camera on {\em ROSAT}.} values from pointed observations.  The RASS count rates always are cited in the instrumental 0.1--2.4~keV band, as are some of the pointed PSPC observations.  However, the so-called WGACAT\footnote{see: http://heasarc.gsfc.nasa.gov/wgacat/.} provides PSPC count rates in a more restricted, albeit more reliable, range of instrumental pulse-height channels, corresponding to 0.24--2.0~keV.  The {\em ROSAT}\/ HRI has virtually no energy discrimination, but its soft response is well-matched to the reference output energy band.  The {\em XMM-Newton}\/ catalogs generally cite observed fluxes in specific energy bandpasses, e.g., the full-band ``ep\_8'' flux (0.2--12~keV).  Such fluxes require an additional conversion factor to translate to the 0.2--2~keV reference band.  In some favorable cases, measurements are available from multiple X-ray observatories, multiple instruments within the same facility (e.g., {\em ROSAT}\/ PSPC and HRI), or from multiple epochs for the same instrument.  In such cases, judgement must be exercised to decide how to combine the multiple measurements, or at what level to discard values that are discrepant compared with the others.  Here, all else being equal, a geometrical average ($n$-th root of product of $n$ values) was applied to the fluxes deemed most reliable.

Specifically for {\em ROSAT,}\/ with the lion's share of measurements, the following ECF relations were adopted (units are erg cm$^{-2}$ count$^{-1}$ for output bandpass 0.2--2~keV):
\begin{equation}
{\rm ECF}= 9{\times}10^{-12} + 3{\times}10^{-11}\,E(B-V) + 1{\times}10^{-12}\,(\log{T}\,-\,7)\,\,,
\end{equation}
for the PSPC and input bandpass 0.24--2~keV (e.g., WGACAT); color excess, $E(B-V)$ (mag); and spectral temperature, $T$ (K).  The latter was taken as $10^{7}$~K for spectrally hard sources, like young solar-type stars, and $10^{6.3}$~K for softer spectra, such as from low-activity dwarfs like $\tau$~Ceti.  Similarly,
\begin{equation}
{\rm ECF}= 7{\times}10^{-12} + 6{\times}10^{-11}\,E(B-V) + 4{\times}10^{-12}\,(\log{T}\,-\,7)\,\,,
\end{equation}
for the PSPC and input bandpass 0.1--2.4~keV; and,
\begin{equation}
{\rm ECF}= 2.2{\times}10^{-11} + 9{\times}10^{-11}\,E(B-V) - 5{\times}10^{-12}\,(\log{T}\,-\,7)\,\,,
\end{equation}
for the HRI (0.1--2.4~keV input bandpass).  These relations were based on WebPIMMS simulations for solar-abundance APEC thermal models and a standard galactic conversion between color excess and hydrogen column, $N_{\rm H}\sim 6{\times}10^{21}\,E(B-V)$~cm$^{-2}$ (e.g., Gudennavar et al.\ 2012).

For the case of {\em Chandra}\/ HRC-I, ECFs were derived for the specific circumstances of each object (e.g., reddening and spectral hardness [if known]) and applied to the literature values of the count rates, accounting for the encircled energy factor if necessary.

The derived X-ray fluxes, and their origins, are summarized in Table~B2.  Note that these fluxes have explicitly been corrected for reddening (by the derived ECFs).  No error bars are cited for the individual fluxes because the (usually small) photometric errors of the snapshot fluxes are irrelevant compared to the much larger systematic uncertainties owing to source variability, as described earlier.

\subsection{FUV}

For all but one case, the FUV fluxes reported in Table~B2 were derived from relatively high resolution {\em HST}\/ archival material, from the COS and STIS spectrographs ($\lambda/\Delta\lambda\sim 2-4{\times}10^{4}$).  The one exception was Alpha Per G dwarf HE\,699, for which the FUV fluxes are published values from the low-dispersion mode of the Goddard High-Resolution Spectrograph (see Table~B2).  The high resolution of the majority of the underlying spectra is important for rejecting blends, and suppressing the influence of the continuum background in these generally warm objects (as least for the G dwarfs and the F and G supergiants).  

Key emission lines were measured in each spectrum by an automated procedure.  Initially, the wavelength scale of the target spectrum was aligned to a pre-selected stellar feature, normally \ion{O}{1}] 1355~\AA\ in the supergiants and \ion{Si}{4} 1394~\AA\ in the dwarfs (where \ion{O}{1}] usually is much fainter than \ion{Si}{4}).  Next, a global continuum fit was derived by applying various filters to the spectrum to suppress the fine-scale emission and absorption structure, leaving ideally a smooth, structure-free background level.  Finally, the target features were numerically integrated in a band specified by a central wavelength, $\lambda_0$, and half-width, $\Delta\lambda$, restricting the integration to the positive fluxes above the derived continuum level (to avoid the bias of deep interstellar absorptions below the continuum level in a few of the target features in the warmer supergiants, although not relevant to the two main diagnostics described earlier, \ion{Si}{4} and \ion{O}{1}]).

The target features, and the parameters specifying the numerical integrations, are listed in Table~B1.  Examples of the fitting procedure are illustrated in Figures~B-1a,-1b, and-1c for representative objects.  The measured fluxes are listed in Table~B2.  There were more features measured than explicitly needed for the flux-flux diagrams presented earlier, and these extra fluxes are included in case they might be of value to other studies.  No error bars are provided for the fluxes, because a numerical integration generally has a very small formal photometric uncertainty owing to the large number of samples included in the bandpass, and the typically high S/N of the underlying spectra.  Further, as with the X-rays noted earlier, the main sources of uncertainty for the FUV fluxes are systematic:  continuum placement, intrinsic variability from epoch to epoch, and uncertain reddening corrections (primarily for the supergiants).  Note that the FUV fluxes in Table~B2 are ``observed'' values (i.e., not corrected for reddening).

There are three Cepheid variables included in the comparison sample, which deserve special mention.  These objects have similar luminosities and effective temperatures to $\alpha$~Per, which is reason enough to include them.  Additional motivation is the remarkable X-ray and FUV behavior described by Engle (2015) and collaborators.  The several Cepheids that have adequate phase-resolved FUV time series display strongly enhanced emission lines at the so-called ``piston phase,'' near maximum optical brightness (but minimum radius).  However, the two Cepheids with time-resolved X-ray measurements -- $\delta$~Cep and $\beta$~Dor -- display sharp high-energy enhancements at an entirely different phase, near that of maximum expansion; but low-level X-ray emission for the rest of the pulsational cycle (including during the FUV high-state).  Because of the anti-correlation between the X-ray and \ion{Si}{4} emissions, the Cepheid X-rays would not seem to align with a traditional coronal explanation (where X-rays and \ion{Si}{4} are strongly associated).  Nevertheless, the X-ray high states of the Cepheids, combined with the respective FUV low-states, mimic in some respects the anomalous $L_{\rm Si\,IV}/L_{\rm X}$ ratios of $\alpha$~Per (see Fig.~6), so identifying the emission mechanism in one case might help with the other.  Thus, further phase-resolved X-ray and FUV measurements of the Cepheid class should be encouraged.

\clearpage
\begin{deluxetable}{lccccl}
\tablenum{B1}
\tablecaption{FUV Features and Flux Integration Limits}
\tablecolumns{6}
\tablewidth{0pt}
\tablehead{\colhead{Transition} & \colhead{$T_{\rm form}$} & \colhead{$\lambda_0$} & \colhead{$\Delta\lambda_{\rm V}$} & 
\colhead{$\Delta\lambda_{\rm I}$}  & \colhead{Notes}  \\[3mm]
\colhead{\AA} & \colhead{(K)} & \colhead{(\AA)} & \colhead{(\AA)} & \colhead{(\AA)} \\[3mm]
\colhead{(1)} & \colhead{(2)} & \colhead{(3)} & \colhead{(4)} & \colhead{(5)}  & \colhead{(6)}  
} 
\startdata
\ion{Si}{3}~1206  & $6{\times}10^4$  &  1206.5  &  0.8  &  1.3  &  \\[3mm]
\ion{N}{5}~1238   & $2{\times}10^5$  &  1238.8  &  0.7  &  1.3  &  \\[3mm]
\ion{N}{5}~1242   & $2{\times}10^5$  &  1242.8  &  0.5  &  1.3  &  \\[3mm]
\ion{O}{1}~1305   & $1{\times}10^4$  &  1305.5  &  1.0  &  1.25  & blend (1304\,+\,1306)  \\[3mm] 
\ion{C}{2}~1335   & $3{\times}10^4$  &  1335.1  &  1.3  &  1.8  & blend (1334\,+\,1335) \\[3mm]
\ion{O}{1}]~1355  & $1{\times}10^4$  &  1355.6  &  0.2\tablenotemark{a}   &  0.4  &  \\[3mm]
\ion{Si}{4}~1394  & $8{\times}10^4$  &  1393.8  &  0.8  &  1.3  &  \\[3mm]
\ion{Si}{4}~1402  & $8{\times}10^4$  &  1402.8  &  0.7  &  1.0  &  \\[3mm]
\enddata
\tablenotetext{a}{Excludes weak \ion{C}{1} blend, just to red of \ion{O}{1}], which tends to be more prominent in dwarfs than supergiants.}
\tablecomments{Col.~2 is the approximate formation temperature from the CHIANTI atomic database (see: http://www.chiantidatabase.org/) for the ions, and a (maximum) chromospheric temperature for the neutrals.  Col.~3 is the central wavelength of the integration bandpass.  Col.~4 is the half width, different for the dwarfs (``V'') and supergiants (``I'').  Total integration band is $(\lambda_0 - \Delta\lambda)~{\rm to}~(\lambda_0 + \Delta\lambda)$.  Col.~6 Notes: ``blend'' indicates that the integration band contains close multiple components of the same species. 
}
\end{deluxetable}

\clearpage
\begin{deluxetable}{lccccccccl}
\rotate
\tablenum{B2}
\tablecaption{X-ray and Far-Ultraviolet Fluxes}
\tablecolumns{10}
\tablewidth{0pt}
\tablehead{\colhead{Name} & \colhead{HD~No.} & \colhead{$f_{\rm X}$} & 
 \colhead{$f_{1206}$} & \colhead{$f_{1240}$\tablenotemark{a}} & \colhead{$f_{1305}$} & \colhead{$f_{1335}$} & 
\colhead{$f_{1355}$} & \colhead{$f_{1400}$\tablenotemark{a}} & \colhead{Notes}  \\[3mm]
\colhead{(1)} & \colhead{(2)} & \colhead{(3)} & \colhead{(4)} & \colhead{(5)}  & \colhead{(6)}  & 
\colhead{(7)} & \colhead{(8)} & \colhead{(9)} & \colhead{(10)}  
} 
\startdata
 $\alpha$~Car        &   45348 & 37 &  5.10 & 3.89 & 8.8 & 2.60: & 3.53 & (5.2)\tablenotemark{b} & Ros; StarC    \\[3mm]
 $\alpha$~Per        &   20902 & 1.7 &  0.074 & 0.075 & 0.29 & 0.034: & 0.059 & 0.085    &  CXO; COS \\[3mm]
 $\delta$~Cep (FUV High-State)\tablenotemark{c}  &  
                        213306 & 0.057 &  0.265: & 0.061 & 1.05 & 0.237: & 0.132 & 0.224    & XMM~[1]; COS      \\[3mm]
 $\delta$~Cep (FUV Low-State)\tablenotemark{d}  &  
                        213306 & 0.057 &  0.020 & 0.016 & 0.345 & 0.016: & 0.030 & 0.023   &  XMM~[1]; COS     \\[3mm]
 $\delta$~Cep (X-ray High-State)\tablenotemark{d}  &  
                        213306 & 0.16 &  0.020 & 0.016 & 0.345 & 0.016: & 0.030 & 0.023   &  XMM~[1]; COS     \\[3mm]
 $\alpha$~UMi         &   8890 & 0.37 &  0.262 & 0.126 & 0.87 & 0.195: & 0.182 & 0.235    &   XMM~[1]; COS    \\[3mm]
 $\beta$~Dor (FUV High-State)\tablenotemark{e}  &   
                         37350 & 0.069 &  0.81: & 0.245 & 4.89: & 0.572: & 0.81 & 0.81    & XMM~[1]; COS     \\[3mm]
 $\beta$~Dor (FUV Low-State)\tablenotemark{f}   &   
                         37350 & 0.069 &  0.089: & 0.065 & 2.17: & 0.070: & 0.100 & 0.116    &  XMM~[1]; COS    \\[3mm]
 $\beta$~Dor (X-ray High-State)\tablenotemark{f}   &   
                         37350 & 0.14 &  0.089: & 0.065 & 2.17: & 0.070: & 0.100 & 0.116    &  XMM~[1]; COS    \\[3mm]
 $\beta$~Aqr         &  204867 & 0.14 &  1.87: & 1.30 & 9.8: & 1.49: & 1.56 & 2.21                 & CXO~[2]; StarC     \\[3mm]
 $\beta$~Cam         &   31910 & 7.9  & 2.85: & 3.37 & 8.3: & 2.30: & 0.99 & 3.20                   & Ros; StarC     \\[3mm]
 $\beta$~Dra         &  159181 & 41 &  12.7: & 9.6 & 38: & 12.4: & 4.80 & 13.5                         & Ros, XMM; ASTR     \\[3mm]
 $\alpha$~Aqr        &  209750 & 0.50 &  3.64: & 2.69 & 17.3: & 2.56: & 2.50 & 4.69                &  CXO~[2]; ASTR  \\[3mm]
 $\zeta$~Dor         &   33262 & 94 &  3.14 & 0.42 & 1.13 & 2.84 & 0.060 & 2.97               &    Ros; StarC   \\[3mm]
 $\chi$~Her          &  142373 & 0.23  & 0.145 & 0.021: & 0.253 & 0.216 & 0.033 & 0.109         & Ros; StarC    \\[3mm]
 $\chi^1$~Ori        &   39587 & 103 &  3.86 & 0.54 & 1.54 & 3.92 & 0.084 & 3.40               &  Ros, XMM; StarC     \\[3mm]
 EK~Dra              &  129333 & 69 &  0.70 & 0.25 & 0.27 & 1.05 & 0.013 & 0.76            &   Ros, XMM; STIS   \\[3mm]
 $\pi^1$~UMa         &   72905 & 55 &  1.86 & 0.36 & 0.77 & 2.20 & 0.037 & 1.83            &   Ros, XMM; STIS   \\[3mm]
 HII\,314            & \nodata & 7.9 &  0.053 & 0.024 & 0.024 & 0.087 & 0.0013 & 0.055     & Ros, XMM; COS       \\[3mm]
 Sun at 1\,pc        & \nodata & 48 &  29 & 6.5 & 26 & 48 & 3.0 & 22                         &  ${\langle}L_{\rm X}/L_{\rm bol}{\rangle}= 1.5{\times}10^{-7}$ [4]; scaled $\alpha$~Cen     \\[3mm]
 HE\,699             & \nodata & 3.6 &  0.033 & 0.020 & \nodata & 0.073 & \nodata & 0.061 &  Ros; GHRS~[3]    \\[3mm]
 $\alpha$~Cen~A      &  128620 & 19 &  19.0 & 4.34 & 17.4 & 32.2 & 2.00 & 14.9                         & ${\langle}L_{\rm X}/L_{\rm bol}{\rangle}= 0.7{\times}10^{-7}$ [4]; ASTR     \\[3mm]
 61~Vir              &  115617 & 0.52 &  0.15 & 0.04: & 0.18 & 0.25 & 0.020 & 0.13         &  Ros; StarC  \\[3mm]
 $\kappa^1$~Cet      &   20630 & 66 &  2.11 & 0.416 & 1.01 & 2.24 & 0.050 & 1.93                & Ros, XMM; StarC  \\[3mm]
 $\xi$~Boo~A         &  131156A & 157 &  3.12 & 0.75 & 1.56 & 4.48 & 0.066 & 3.13           & Ros, XMM; StarC      \\[3mm]
 $\tau$~Cet          &   10700 &  2.1 &  0.355 & 0.053 & 0.76 & 0.68 & 0.080 & 0.280         &  Ros; StarC     \\[3mm]
 HR\,8               &     166 &  50 &  1.00 & 0.32 & 0.53 & 1.24 & 0.030 & 0.97              &   Ros; StarC    \\[3mm]
 70~Oph~A            &  165341A &  65 &  2.33 & 0.62 & 1.85 & 3.98 & 0.098 & 1.98            &  Ros, XMM; StarC     \\[3mm]
 $\alpha$~Cen~B      &  128621 &  64 &  13.2 & 3.98 & 13.8 & 26.0 & 1.04 & 11.4                 & ${\langle}L_{\rm X}/L_{\rm bol}{\rangle}= 7{\times}10^{-7}$ [4]; ASTR     \\[3mm]
 $\epsilon$~Eri          &   22049 &  150 &  3.51 & 1.39 & 3.59 & 6.40 & 0.183 & 3.38       &  Ros, XMM; StarC     \\[3mm]
 AU~Mic              &  197481 & 250 &  1.36 & 1.37 & 0.78 & 2.25 & 0.038 & 1.70            &  Ros, XMM; StarC     \\[3mm]
 AD~Leo              & \nodata & 180 &  1.31 & 1.33 & 0.77 & 2.18 & 0.036 & 1.59            &  Ros, XMM; StarC   \\[3mm]
\enddata
\tablenotetext{a}{Combined flux from individual component measurements: see Table~B1.}
\tablenotetext{b}{Scaled from Si~{\scriptsize III} assuming de-reddened fluxes are 1:1 (directly measured value 10.1).}
\tablenotetext{c}{COS datasets lbk8:\,15010+20010}
\tablenotetext{d}{COS datasets lbk8:\,09010+17010+18010+23010}
\tablenotetext{e}{COS datasets lbk8:\,01010+10010+11010}
\tablenotetext{f}{COS dataset lbk814010}
\tablecomments{Col.~3--9 fluxes in $10^{-13}$ erg cm$^{-2}$ s$^{-1}$ at Earth (no reddening correction for Cols.~4--9).  Col.~3 X-ray energy band: 0.2--2~keV.  Colons indicate uncertain values, due to weak feature, bright continuum, or spectral complexity (e.g., strong interstellar and/or circumstellar absorption). ``Sun at 1\,pc'' values were scaled from $\alpha$~Cen~A according to ratios with cycle-average solar FUV irradiance spectra from SUSIM (http://wwwsolar.nrl.navy.mil/susim\_uars.html).  The ratios were derived using stellar spectra smoothed to the 1.5~\AA\ resolution of the highest dispersion SUSIM data products.  In the Col.~10 Notes, the source of the X-ray flux is listed first, then the FUV, separated by a semicolon.  The abbreviations are: Ros= {\em ROSAT}\/ PSPC and/or HRI; CXO= {\em Chandra}\/ HRC-I; XMM= {\em XMM-Newton}\/ EPIC; StarC= StarCAT
(https://archive.stsci.edu/prepds/starcat/); ASTR= ASTRAL (http://casa.colorado.edu/$\sim$ayres/ASTRAL/).\\[3mm]
{\em References.}\/ [1] Engle (2015): $2-T$ values; [2] Ayres et al.\ (2005); [3] Ayres (1999); [4] Ayres (2015b).}
\end{deluxetable}

\clearpage
\begin{figure}[ht]
\figurenum{B-1a}
\vskip  -5mm
\hskip  8mm
\includegraphics[scale=0.85]{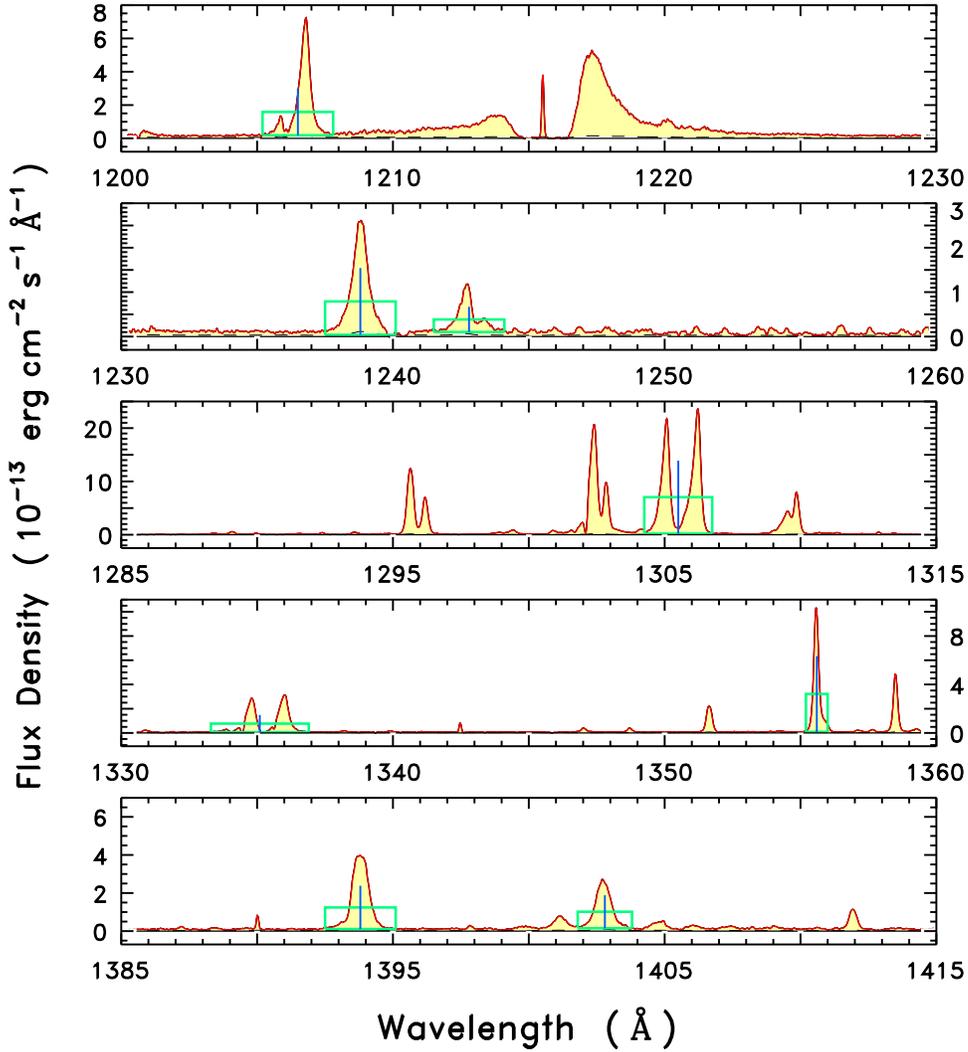} 
\vskip  0mm
\figcaption[]{\small  Schematic rendition of automated line flux integrations.  The FUV spectrum is for $\alpha$~Aqr (STIS medium-resolution FUV echelle, from the ASTRAL catalog): observed fluxes, no reddening correction.  The narrow spike in the center of the broad Ly$\alpha$ 1215~\AA\ interstellar absorption trough is geocoronal \ion{H}{1} emission.  Other narrow features (e.g., 1390~\AA) are hot pixels.  The black dashed curve (barely visible) is the smoothed 1\,$\sigma$ photometric noise (per resel).  The wavelength scale was registered to the apparent velocity of the \ion{O}{1}] 1355~\AA\ intersystem transition, which is sharp and unaffected by circumstellar or interstellar absorptions (unlike the resonance triplet lines at 1305~\AA).  Blue ticks mark the central wavelength of the integration bandpass, $\lambda_0$, while the sides of the green boxes show the extent (${\pm}\Delta\lambda$).  The lower boundary of the box is the derived continuum level, while the upper boundary is the average flux density of the integrated feature (area of box is the integrated flux).  The features measured in the spectrum are listed in Table~B1.  The supergiant integration template was used for this broad-line star.
}
\end{figure}

\clearpage
\begin{figure}[ht]
\figurenum{B-1b}
\vskip  -5mm
\hskip  8mm
\includegraphics[scale=0.85]{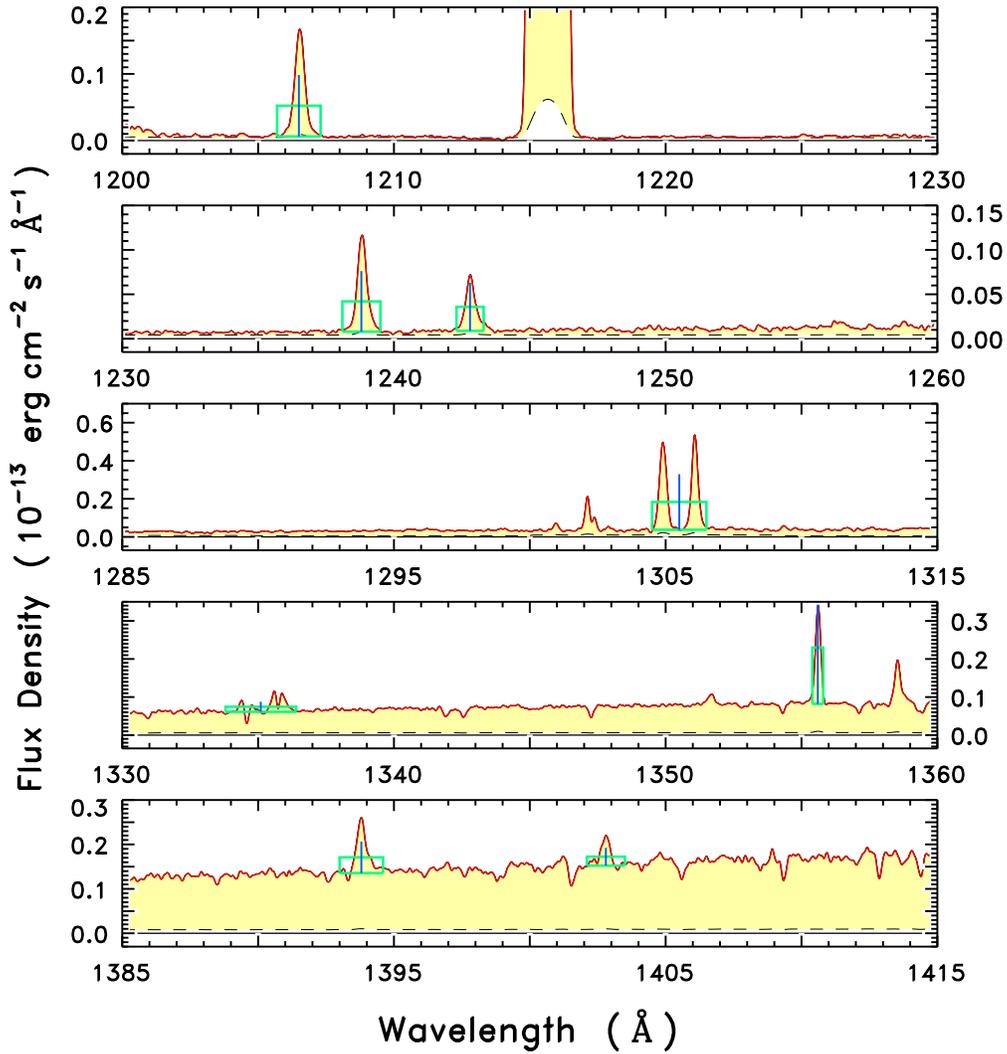} 
\vskip  0mm
\figcaption[]{\small  Same as Fig.~B-1a for $\alpha$~Per (COS G130M, from this study).  The bright feature at Ly$\alpha$ is overexposed geocoronal emission through the 2.5$\arcsec$-diameter Primary Science Aperture.  The dwarf integration template was used, because the FUV lines of the supergiant are atypically narrow for its class.
}
\end{figure}

\clearpage
\begin{figure}[h]
\figurenum{B-1c}
\vskip  -5mm
\hskip  8mm
\includegraphics[scale=0.85]{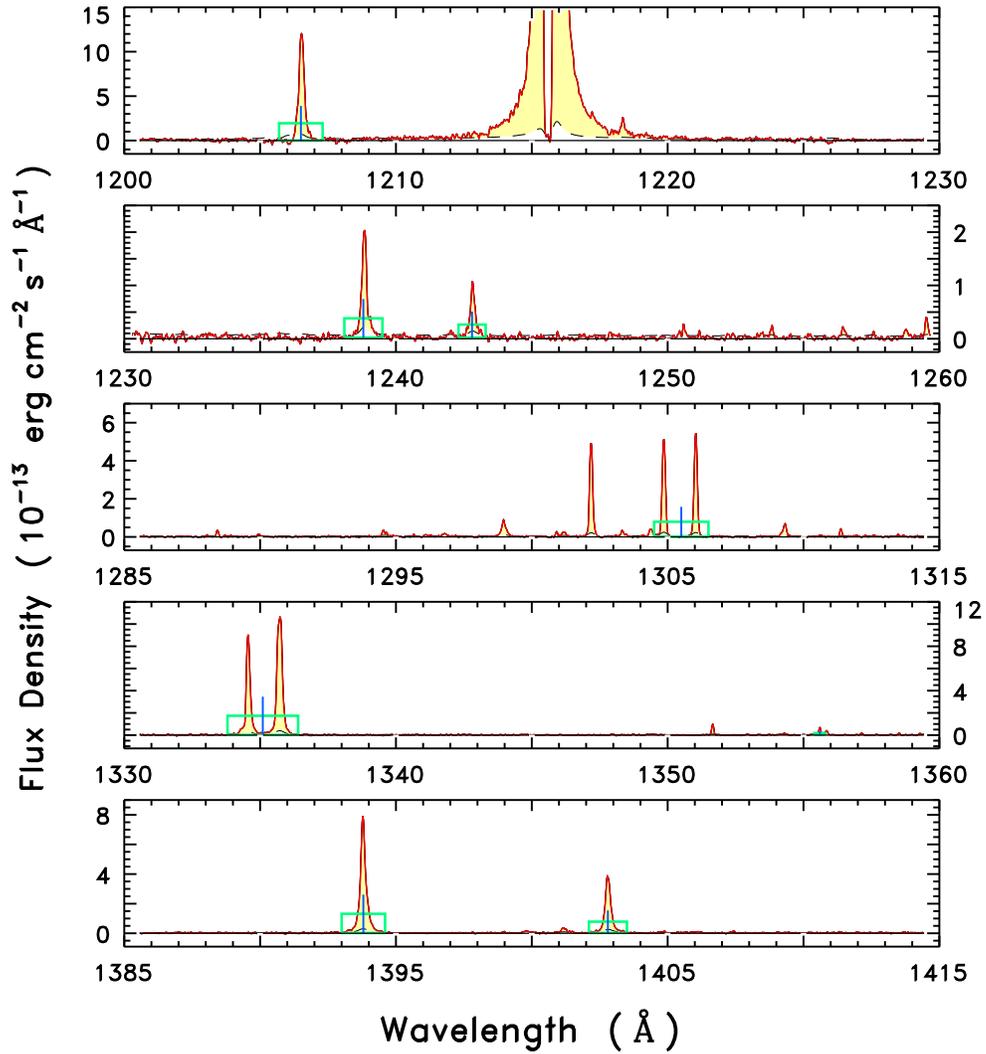} 
\vskip  0mm
\figcaption[]{\small  Same as Fig.~B-1a for the late-G dwarf $\xi$~Boo~A (STIS medium-resolution FUV echelle, from StarCAT).  In this case, the wavelength registration was according to \ion{Si}{4} 1393~\AA, owing to the faintness of the normal calibrator, \ion{O}{1}] 1355~\AA.  The dwarf integration template was used.
}
\end{figure}

\end{document}